\documentclass[12pt,preprint]{aastex}
\usepackage{lineno}
\usepackage{url}

\newcommand{\Fermic}{\textit{Fermi}}
\newcommand{\Fermi}{\Fermic\ }
\newcommand{\FermiLATc}{\Fermic-LAT}
\newcommand{\FermiLAT}{\FermiLATc\ }

\newcommand{\Swiftc}{\textit{Swift}}
\newcommand{\Swift}{\Swiftc\ }

\newcommand{\RXTEc}{RXTE}
\newcommand{\RXTE}{\RXTEc\ }

\newcommand{\etal}{\MakeLowercase{\textit{et al. }}} % "et al."

\newcommand{\lapp}{\ensuremath{\stackrel{\scriptstyle <}{{}_{\sim}}}}
\newcommand{\gapp}{\ensuremath{\stackrel{\scriptstyle >}{{}_{\sim}}}}

\def\us{\char`\_}

\slugcomment{Submitted to Astrophysical Journal}

\shorttitle{Fermi view of Mrk~421}
\shortauthors{Abdo et al.}

\begin{document}


\title{\FermiLAT  Observations of Markarian 421: the Missing Piece of its Spectral Energy Distribution}


\author{
A.~A.~Abdo\altaffilmark{2}, 
M.~Ackermann\altaffilmark{3}, 
M.~Ajello\altaffilmark{3}, 
L.~Baldini\altaffilmark{4}, 
J.~Ballet\altaffilmark{5}, 
G.~Barbiellini\altaffilmark{6,7}, 
D.~Bastieri\altaffilmark{8,9}, 
K.~Bechtol\altaffilmark{3}, 
R.~Bellazzini\altaffilmark{4}, 
B.~Berenji\altaffilmark{3}
% ...
\\
(The \FermiLAT collaboration) \\
J.~Aleksi\'c\altaffilmark{67}, 
L.~A.~Antonelli\altaffilmark{68}, 
P.~Antoranz\altaffilmark{69}, 
M.~Backes\altaffilmark{70}, 
J.~A.~Barrio\altaffilmark{71}, 
J.~Becerra~Gonz\'alez\altaffilmark{72,73}, 
W.~Bednarek\altaffilmark{74}, 
A.~Berdyugin\altaffilmark{75}, 
K.~Berger\altaffilmark{73}, 
E.~Bernardini\altaffilmark{76}
% ...
\\
(The MAGIC collaboration) \\ 
M.~Villata\altaffilmark{110}, 
C.~Raiteri\altaffilmark{110}, 
H.~D.~Aller\altaffilmark{92}, 
M.~F.~Aller\altaffilmark{92}, 
W.~P.~Chen\altaffilmark{95}, 
B.~Jordan\altaffilmark{99}, 
E.~Koptelova\altaffilmark{95}, 
O.~M.~Kurtanidze\altaffilmark{101}, 
A.~L\"ahteenm\"aki\altaffilmark{98}, 
B.~McBreen\altaffilmark{18}, 
V.~M.~Larionov\altaffilmark{102,103,104}, 
C.~S.~Lin\altaffilmark{95}, 
M.~G.~Nikolashvili\altaffilmark{101}, 
R.~Reinthal\altaffilmark{75}, 
E.~Angelakis\altaffilmark{33}, 
M.~Capalbi\altaffilmark{20}, 
A.~Carrami\~nana\altaffilmark{93}, 
L.~Carrasco\altaffilmark{93},
P.~Cassaro\altaffilmark{105},
A.~Cesarini\altaffilmark{94}, 
A.~Falcone\altaffilmark{96}, 
M.~A.~Gurwell\altaffilmark{97}, 
T.~Hovatta\altaffilmark{98}, 
Yu.~A.~Kovalev\altaffilmark{100},
Y.~Y.~Kovalev\altaffilmark{100,33}, 
T.~P.~Krichbaum\altaffilmark{33}, 
H.~A.~Krimm\altaffilmark{21,42}, 
%P.~Leto\altaffilmark{105}, 
M.~L.~Lister\altaffilmark{106}, 
J.~W.~Moody\altaffilmark{107}, 
G.~Maccaferri\altaffilmark{114}, 
Y.~Mori\altaffilmark{64}, 
I.~Nestoras\altaffilmark{33}, 
A.~Orlati\altaffilmark{114}, 
C.~Pace\altaffilmark{107}, 
C.~Pagani\altaffilmark{108}, 
R.~Pearson\altaffilmark{107}, 
M.~Perri\altaffilmark{20}, 
B.~G.~Piner\altaffilmark{109}, 
E.~Ros\altaffilmark{33,111}, 
A.~C.~Sadun\altaffilmark{112}, 
T.~Sakamoto\altaffilmark{17}, 
J.~Tammi\altaffilmark{98}, 
A.~Zook\altaffilmark{113}
}
% ...
\altaffiltext{2}{National Research Council Research Associate, National Academy of Sciences, Washington, DC 20001, resident at Naval Research Laboratory, Washington, DC 20375, USA}
\altaffiltext{3}{W. W. Hansen Experimental Physics Laboratory, Kavli Institute for Particle Astrophysics and Cosmology, Department of Physics and SLAC National Accelerator Laboratory, Stanford University, Stanford, CA 94305, USA}
\altaffiltext{4}{Istituto Nazionale di Fisica Nucleare, Sezione di Pisa, I-56127 Pisa, Italy}
\altaffiltext{5}{Laboratoire AIM, CEA-IRFU/CNRS/Universit\'e Paris Diderot, Service d'Astrophysique, CEA Saclay, 91191 Gif sur Yvette, France}
\altaffiltext{6}{Istituto Nazionale di Fisica Nucleare, Sezione di Trieste, I-34127 Trieste, Italy}
\altaffiltext{7}{Dipartimento di Fisica, Universit\`a di Trieste, I-34127 Trieste, Italy}
\altaffiltext{8}{Istituto Nazionale di Fisica Nucleare, Sezione di Padova, I-35131 Padova, Italy}
\altaffiltext{9}{Dipartimento di Fisica ``G. Galilei", Universit\`a di Padova, I-35131 Padova, Italy}
% ...
\altaffiltext{17}{NASA Goddard Space Flight Center, Greenbelt, MD 20771, USA}
\altaffiltext{18}{University College Dublin, Belfield, Dublin 4, Ireland}
% ...
\altaffiltext{20}{Agenzia Spaziale Italiana (ASI) Science Data Center, I-00044 Frascati (Roma), Italy}
\altaffiltext{21}{Center for Research and Exploration in Space Science and Technology (CRESST) and NASA Goddard Space Flight Center, Greenbelt, MD 20771, USA}
% ...
\altaffiltext{33}{Max-Planck-Institut f\"ur Radioastronomie, Auf dem H\"ugel 69, 53121 Bonn, Germany}
% ...
\altaffiltext{42}{Universities Space Research Association (USRA), Columbia, MD 21044, USA}
% ...
\altaffiltext{64}{Department of Physics, Tokyo Institute of Technology, Meguro City, Tokyo 152-8551, Japan}
% ...
\altaffiltext{67}{Institut de F\'isica d'Altes Energies (IFAE), Edifici Cn, Universitat Aut\`onoma de Barcelona (UAB), E-08193 Bellaterra (Barcelona), Spain}
\altaffiltext{68}{INAF National Institute for Astrophysics, I-00136 Roma, Italy}
\altaffiltext{69}{Universit\`a di Siena and INFN Pisa, I-53100 Siena, Italy}
\altaffiltext{70}{Technische Universit\"at Dortmund, D-44221 Dortmund, Germany}
\altaffiltext{71}{Universidad Complutense, E-28040 Madrid, Spain}
\altaffiltext{72}{Instituto de Astrof\'isica de Canarias, E38205 - La Laguna (Tenerife), Spain}
\altaffiltext{73}{Departamento de Astrofisica, Universidad de La Laguna, E-38205 La Laguna, Tenerife, Spain}
\altaffiltext{74}{University of  {\L}\'od\'z, PL-90236 {\L}\'od\'z, Poland}
\altaffiltext{75}{Tuorla Observatory, University of Turku, FI-21500 Piikki\"o, Finland}
\altaffiltext{76}{Deutsches Elektronen Synchrotron DESY, D-15738 Zeuthen, Germany}
% ...
\altaffiltext{92}{Department of Astronomy, University of Michigan, Ann Arbor, MI 48109-1042, USA}
\altaffiltext{93}{Instituto Nacional de Astrof\'isica, \'Optica y Electr\'onica, Tonantzintla, Puebla 72840, Mexico}
\altaffiltext{94}{Physics Department, National University of Ireland Galway, Ireland}
\altaffiltext{95}{Graduate Institute of Astronomy, National Central University, Jhongli 32054, Taiwan}
\altaffiltext{96}{Department of Astronomy and Astrophysics, Pennsylvania State University, University Park, PA 16802, USA}
\altaffiltext{97}{Harvard-Smithsonian Center for Astrophysics, Cambridge, MA 02138, USA}
\altaffiltext{98}{Aalto University Mets\"ahovi Radio Observatory, FIN-02540 Kylmala, Finland}
\altaffiltext{99}{School of Cosmic Physics, Dublin Institute for Advanced Studies, Dublin, 2, Ireland}
\altaffiltext{100}{Astro Space Center of the Lebedev Physical Institute, 117997 Moscow, Russia}
\altaffiltext{101}{Abastumani Observatory, Mt. Kanobili, 0301 Abastumani, Georgia}
\altaffiltext{102}{Isaac Newton Institute of Chile, St. Petersburg Branch, St. Petersburg, Russia}
\altaffiltext{103}{Pulkovo Observatory, 196140 St. Petersburg, Russia}
\altaffiltext{104}{Astronomical Institute, St. Petersburg State University, St. Petersburg, Russia}
\altaffiltext{105}{INAF Istituto di Radioastronomia, Sezione di Noto,Contrada Renna Bassa, 96017 Noto (SR), Italy}
%\altaffiltext{105}{Osservatorio Astrofisico di Catania, 95123 Catania, Italy}
\altaffiltext{106}{Department of Physics, Purdue University, West Lafayette, IN 47907, USA}
\altaffiltext{107}{Department of Physics and Astronomy, Brigham Young University, Provo, Utah 84602, USA}
\altaffiltext{108}{Department of Physics and Astronomy, University of Leicester, Leicester, LE1 7RH, UK}
\altaffiltext{109}{Department of Physics and Astronomy, Whittier College, Whittier, CA, USA}
\altaffiltext{110}{INAF, Osservatorio Astronomico di Torino, I-10025 Pino Torinese (TO), Italy}
\altaffiltext{111}{Universitat de Val\`encia, 46010 Val\`encia, Spain}
\altaffiltext{112}{Department of Physics, University of Colorado, Denver, CO 80220, USA}
\altaffiltext{113}{Department of Physics and Astronomy, Pomona College, Claremont CA 91711-6312, USA}
\altaffiltext{114}{INAF Istituto di Radioastronomia, Stazione Radioastronomica di Medicina, I-40059 Medicina (Bologna), Italy}

\begin{abstract}

We report on the $\gamma$-ray activity of the high-synchrotron-peaked
BL Lacertae object Mrk\,421 during the first 1.5 years of \Fermi
operation, from 2008 August 5 to 2010 March 12. We find that the Large
Area Telescope (LAT) $\gamma$-ray spectrum above $0.3$\,GeV can be
well-described by a power-law function with photon index $\Gamma=1.78
\pm 0.02$ and average photon flux $F(>0.3$\,GeV$)=(7.23 \pm 0.16)
\times 10^{-8}$\,ph\,cm$^{-2}$\,s$^{-1}$. Over this time period, the
\FermiLAT spectrum above $0.3$\,GeV was evaluated on 7-day-long time
intervals, showing significant variations in the photon flux (up to a
factor $\sim 3$ from the minimum to the maximum flux), but mild
spectral variations.  The variability amplitude at X-ray frequencies
measured by \mbox{\RXTEc/ASM} and \Swiftc/BAT is substantially larger
than that in $\gamma$-rays measured by \FermiLATc, and these two
energy ranges are not significantly correlated.  We also present the
first results from the 4.5-month-long multifrequency campaign on
Mrk\,421, which included the VLBA, \Swiftc, \RXTEc, MAGIC, the
F-GAMMA, GASP-WEBT, and other collaborations and instruments which
provided excellent temporal and energy coverage of the source throughout
the entire campaign (2009 January 19 to 2009 June 1). During this
campaign, Mrk\,421 showed a low activity at all wavebands. The
extensive multi-instrument (radio to TeV) data set provides an
unprecedented, complete look at the quiescent spectral energy
distribution (SED) for this source.  The broad band SED was reproduced 
with a leptonic (one-zone Synchrotron Self-Compton) and a hadronic
model (Synchrotron Proton Blazar). Both frameworks are able to
describe the average SED reasonably well, implying comparable jet
powers but very different characteristics for the blazar emission
site.

%In the particular 
%case of the leptonic scenario, the larger-than canonical
%breaks required in the electron energy distribution  to reproduce
%the observed SED suggests departures from a homogenous emission region and
%possibly velocity gradients.

\end{abstract}



%% ============================================================================
%%
%% SECTION 1 -- INTRODUCTION
%%
%% ============================================================================

\section{Introduction}
\label{Intro}

Blazars are active galaxies believed to have pairs of relativistic
jets flowing in opposite directions closely aligned to our line of
sight.  Their spectral energy distributions (SEDs) are dominated by
beamed jet emission and take the form of two broad nonthermal
components, one at low energy, peaking in the radio through optical,
and one at high energies, peaking in the $\gamma$-rays.  Some blazars
have been well-monitored for decades and along a wide range of
wavelengths.  Although there is ample evidence for the electron
synchrotron origin of the low-energy bump, the existing data do not
allow an unambiguous identification of the radiation mechanism
responsible for the high-energy bump.  One reason for this is that the
high-energy bump is poorly constrained due to the lack of observations
at energies between $\sim$ 0.1~MeV and 0.3~TeV.  This gap was filled
to some extent by EGRET on the {\em Compton Gamma-Ray Observatory}
\citep[]{Hartman1999}.  However, its moderate sensitivity and limited
observing time precluded detailed cross-correlation studies between
$\gamma$-ray and lower-energy wavebands. On the other hand, the
current generation of TeV imaging atmospheric Cherenkov telescopes
(IACTs), HESS, MAGIC, and VERITAS, which have good sensitivity
at energies as low as 0.1 TeV, did not start scientific operation until
2004; that is, well after EGRET had stopped operating.

This has changed with the launch of the \Fermi Gamma-ray Space
Telescope in June 2008.  In science operation since 2008 August, its
LAT instrument \citep{Atwood2009} views the entire sky in the 20 MeV to greater than 300
GeV range every three hours.  The one year LAT AGN Catalog
\citep[1LAC;][]{ref22} contains around 600 blazars, a factor of $\sim 10$
greater than EGRET detected during its entire operational
lifetime. For the first time, simultaneous 
observations of \Fermi with the latest  generation of IACTs can cover the entire 
high-energy bump. Combining this with simultaneous low-energy
observations gives an unprecedented multiwavelength view of
these enigmatic objects.

Blazars found in low states are particularly poorly studied.  This is
due in part to the lower sensitivity of previous instruments, and in
part to the fact that multiwavelength monitoring programs, including
space-based instruments, are mostly triggered when an object enters a
particularly bright state, as observed by ground-based optical
telescopes and all-sky monitors such as the {\em RXTE}
\citep[]{RXTERef} All Sky Monitor (ASM) or the {\em Swift}
\citep[]{SwiftRef} Burst Alert Telescope (BAT). Having a well-measured
low-state SED will be useful for constraining models and as a baseline
to which other, flaring states can be compared.  This will be crucial
for answering many of the questions regarding these objects.

Markarian 421 (Mrk\,421; RA=11$^h$ 4$^m$ 27.31$^s$, Dec=38$^\circ$ 12'
31.8" , J2000, redshift $z = 0.031$) is a high-synchrotron-peaked
(HSP) BL Lac (according to the classification presented in
\citet{latsed}), that is one of the brightest sources in the
extragalactic X-ray/TeV sky.  Mrk\,421 was actually the first
extragalactic object to be discovered as a TeV emitter
\citep{Punch1992}, and one of the fastest varying $\gamma$-ray sources
\citep{Gaidos1996}.  During the last decade, there were a large number
of publications on the very high energy (VHE) $\gamma$-ray spectrum of
this source, which has been measured with almost all the existing
IACTs
\citep{Krennrich2002,Aharonian2002,Aharonian2003,Aharonian2005,Mrk421MAGIC,Acciari2009}.
Among other things, we learned that the source shows evidence for a
spectral hardening with increasing flux.  The SED and the
multifrequency correlations of Mrk\,421 have also been intensively
studied in the past through dedicated multifrequency observations of
the source
\citep{Katar2003,Blazejowski2005,Mrk421Whipple2006,fossati08,Horan2009},
which showed a positive but very complex relation between X-rays and
VHE $\gamma$-rays, and that a simple one-zone Synchrotron Self-Compton
model (SSC) with an electron distribution parameterized with one or
two power-laws seemed to describe the collected SED well during the
observing campaigns. During a strong flare in June 2008, the source
was also detected with the gamma-ray telescope AGILE and, for the first time, a hint of
correlation between optical and TeV energies was reported by
\cite{Donnarumma2009}.

Despite the large number of publications on Mrk\,421, the details of
the physical processes underlying the blazar emission are still
unknown. The main reasons for this are the sparse multifrequency data
during long periods of time, and the moderate sensitivity available in
the past to study the $\gamma$-ray emission of this source. In
addition, as occurs often with studies of blazars, many of the
previous multifrequency campaigns were triggered by an enhanced flux
level at X-rays and/or $\gamma$-rays, and hence many of the previous
studies of this source are biased towards ``high-activity'' states,
where perhaps distinct physical processes play a dominant role.
Moreover, we have very little information from the MeV-GeV energy
range: 9 years of operation with EGRET resulted in only a few viewing
periods with a signal significance of barely 5 standard deviations
($\sigma$, hereafter) \citep{Hartman1999}, which precluded detailed
correlation studies with other energy bands.

We took advantage of the new capabilities provided by \FermiLAT and
the new IACTs, as well as the existing capabilities for observing at
X-ray and lower frequencies, and organized a multifrequency (from
radio to TeV) campaign to observe Mrk~421 over 4.5 months. 
%The observing campaign started on 2009 January 19 (MJD 54850) and
%finished on 2009 June 1 (MJD 54983). 
The observational goal for this campaign
was to sample Mrk~421 every 2 days, which was accomplished 
at optical, X-rays and TeV energies whenever the weather and/or
technical operations allowed.  The \FermiLAT operated in survey mode
and thus the source was constantly observed at $\gamma$-ray energies.
%The instruments that participated and the observing schedule of the
%campaign can be found online\footnote{
 % See \url{https://confluence.slac.stanford.edu/display/GLAMCOG/Campaign+on+Mrk421+(Jan+2009+to+May+2009)} \\
 % Mantained by D. Paneque.}.  
In this paper, we report the overall
SED averaged over the duration of the observing campaign. A more in-depth analysis of the
multifrequency data set (variability, correlations and implications)
will be given in a forthcoming paper.

The work is organized as follows: In section 2 we introduce the LAT instrument
and report on the data analysis. In section 3 we report the
flux/spectral variability in the $\gamma$-ray range observed by
\FermiLAT during the first 1.5 years of operation, and compare it
with the flux variability obtained with \RXTEc/ASM and \Swiftc/BAT, which
are also all-sky instruments. In section 4 we report on the spectrum
of Mrk~421 measured by Fermi, and section 5 reports on the overall SED
collected during the the 4.5 month long multi-wavelength campaign
organized in 2009.  Section 6 is devoted to SED modeling of
the multifrequency data with both a hadronic and a leptonic model, 
and in section 7 we discuss the implications
of the experimental and modeling results.  Finally, we conclude on section 8.

%% ============================================================================
%%
%% SECTION 2 -- \FermiLAT data selection and analysis
%%
%% ============================================================================

\section{\FermiLAT Data Selection and Analysis}
\label{FermiData}

The \FermiLAT is a $\gamma$-ray telescope operating from $20$\,MeV to
$>300$\,GeV. The instrument is an array of $4 \times 4$ identical
towers, each one consisting of a tracker (where the photons are
pair-converted) and a calorimeter (where the energies of the
pair-converted photons are measured). The entire instrument is covered
with an anticoincidence detector to reject the charged-particle
background. The LAT has a large peak effective area ($0.8$\,m$^2$ for
$1$\,GeV photons), an energy resolution typically better than $10\%$,
and a field of view (FoV) of about $ 2.4$\,sr with an angular
resolution ($68\%$ containment angle) better than 1$^{\circ}$ for
energies above $1$\,GeV. Further details on the description of LAT
are given by \cite{Atwood2009}.

The LAT data reported in this paper were collected from 2008 August 5
(MJD 54683) to 2010 March 12 (MJD 55248). During this time, the
\FermiLAT instrument operated almost entirely in survey mode.  The
analysis was performed with the ScienceTools software package version
v9r15p6. Only events having the highest probability of being photons,
those in the ``diffuse'' class, were used. The LAT data were extracted
from a circular region with a $10^{\circ}$ radius centered at the
location of Mrk~421. The spectral fits were performed using photon
energies greater than $0.3$ GeV, where the effective area of the
instrument is large ($>0.5~$m$^{2}$) and the angular resolution
relatively good ($68\%$ containment angle smaller than
2$^{\circ}$). The spectral fits using energies above $0.3$\,GeV are
less sensitive to possible contamination from non-accounted
(transient) neighboring sources, and have smaller systematic errors,
at the expense of reducing somewhat the number of photons from the
source. In addition, a cut on the zenith angle ($< 105^{\circ}$) was
also applied to reduce contamination from the Earth limb 
$\gamma$-rays, which are produced by cosmic rays interacting with the
upper atmosphere.

The background model used to extract the $\gamma$-ray signal includes
a Galactic diffuse emission component and an isotropic component.  The
model that we adopted for the Galactic component is given by the file
gll\_iem\_v02.fit, and the isotropic component, which is the sum of
the extragalactic diffuse emission and the residual charged particle
background, is parametrized by the file isotropic\_iem\_v02 \footnote{
\url{http://fermi.gsfc.nasa.gov/ssc/data/access/lat/BackgroundModels.html}}.
The normalization of both components in the background model were
allowed to vary freely during the spectral point fitting. {\bf The spectral analyses (from which we derived spectral fits and photon
fluxes) were performed with the post-launch instrument 
response functions \texttt{P6\_V3\_DIFFUSE} using an unbinned maximum likelihood
method.
The systematic uncertainty in
the flux were estimated as $10\%$ at $0.1$\,GeV, $5\%$ at $560$\,MeV
and $20\%$ at $10$\,GeV and above\footnote{See
\texttt{http://fermi.gsfc.nasa.gov/ssc/data/analysis/LAT\us
caveats.html}}.}

%% ============================================================================
%%
%% SECTION 3 -- Flux and spectral variability 
%%
%% ============================================================================

%\vspace{-0.25cm}
\section{Flux and Spectral Variability}
\label{LC}

\begin{figure}[b]
  \centering
\includegraphics[height=2.2in,width=4.0in]{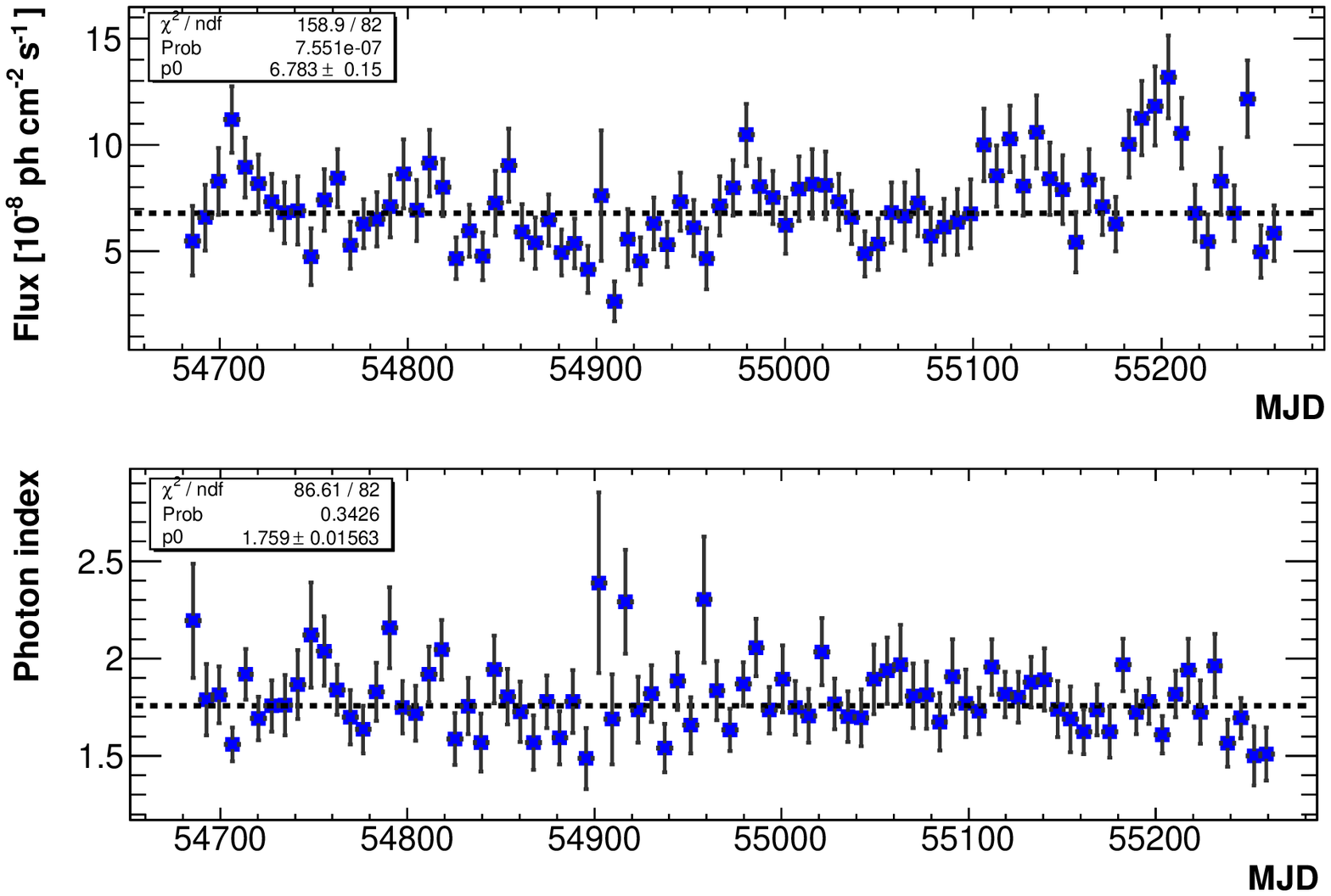}
\includegraphics[height=2.2in,width=2.4in]{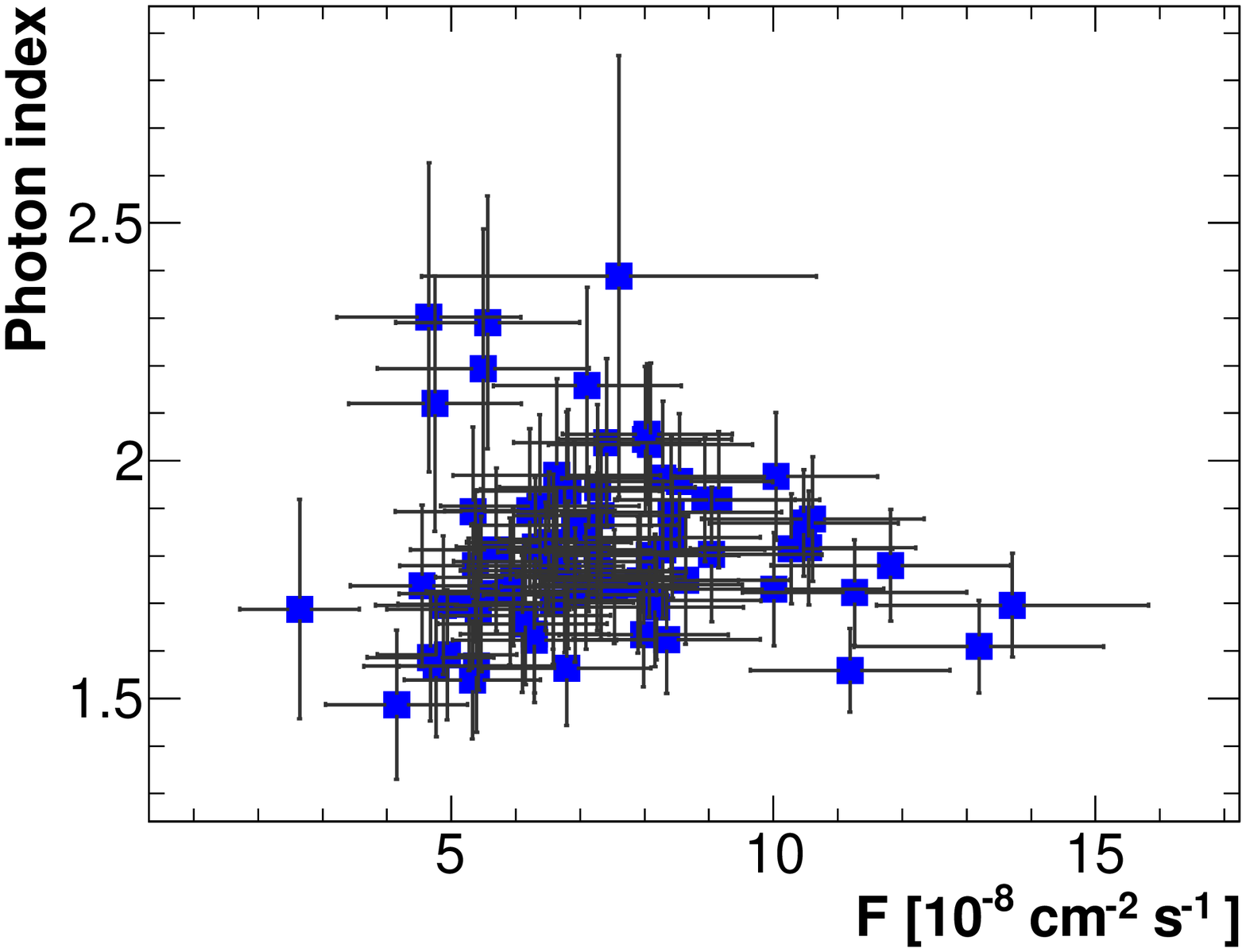}
\caption{ {\bf Left:} $\gamma$-ray flux at photon energies above $0.3$\,GeV (top) and spectral photon index from a power-law fit
    (bottom) for Mrk\,421 for 7-day-long time intervals from 2008
    August 5 (MJD 54683) to 2009 March 12 (MJD 55248). Vertical bars denote $1 \sigma$
    uncertainties and the horizontal error bars denote the width of
    the time interval. The black dashed line and legend show the results from a constant fit to the entire data set. {\bf Right:} scatter plot of the photon index
    versus flux.}
 \label{fig:Lc7days}
 \end{figure}

The sensitivity of \FermiLAT is sufficient to accurately monitor the
$\gamma$-ray flux of Mrk~421 on short timescales (few days)\footnote{
{\bf The number of photons from Mrk\,421 (above 0.3 GeV) detected by LAT in one day is
typically about 6.}}.  The measured $\gamma$-ray flux above 0.3~GeV and the photon index from a
power law (PL) fit are shown in Figure~\ref{fig:Lc7days}.  The data
span the time from 2008 August 5 (MJD 54683) to 2009 March 12 (MJD
55248) and they are binned on time intervals of 7 days. The Test
Statistic (TS) values\footnote{The Test Statistic TS=$2\Delta log(likelihood)$ between models with and without the source is a
  measure of the  probability of having a point $\gamma$-ray source at the location
specified. The TS value is related to the significance of the signal \citep{Mattox1996}.} for the 81
time intervals are typically well in excess of 100 ($\sim$10
sigma). The number of intervals with TS$<$100 is only 9 (11\%). The
lowest TS value is 30, which occurs for the time interval MJD 54899-54906. This low signal significance is due to the fact that the
\FermiLAT instrument did not operate during the time interval MJD
54901-54905\footnote{ The LAT did not operate during the time interval
MJD 54901-54905 due to an unscheduled shutdown.}  and hence only three out of the seven days of the
interval contain data.  The second lowest TS value is 40, which
occurred for the time interval 54962-54969.  During the first 19
months of \Fermi operation, Mrk\,421 showed relatively mild
$\gamma$-ray flux variations, with the lowest photon flux $F(>0.3$ GeV$) = 
(2.6\pm0.9) \times 10^{-8} $cm$^{-2}$ s$^{-1}$ (MJD 54906-54913; TS=53)
and the highest $F(>0.3$ GeV$) =(13.2\pm1.9) \times 10^{-8} $cm$^{-2} $s$^{-1}$
(MJD 55200-55207; TS=355). A constant fit to the flux points from
Figure~\ref{fig:Lc7days} gave a $\chi^2$= 159 for 82 degrees of
freedom (probability the flux was constant is $8 \times 10^{-7}$),
hence indicating the existence of statistically significant flux
variability. On the other hand, the photon index measured in
7-day-long time intervals is statistically compatible with being
constant, as indicated by the results of the constant fit to all the
photon index values, which gave $\chi^2$= 87 for 82 degrees of freedom
(probability no variability is $0.34$). The scatter plot with Flux
versus Index in Figure~\ref{fig:Lc7days} shows that there is no
obvious relation between these two quantities. We quantified the
correlation as prescribed in \citet{Edelson1988}, obtaining a Discrete
Correlation Function $DCF= 0.06
\pm 0.11 $ for a time lag of zero.

 It is interesting to compare the $\gamma$-ray fluxes measured by
 \Fermi with those historical ones recorded by EGRET. From the 3rd
 EGRET catalog \citep{Hartman1999}, one can see that the
 highest/lowest significantly-measured ($TS>25$) photon fluxes are
 $F^{Max} (>0.1$ GeV$) = (27.1\pm6.9) \times 10^{-8} $cm$^{-2}$ s$^{-1}$ (TS=32)
 and $F^{Min} (>0.1$ GeV$) = (10.9\pm2.8) \times 10^{-8}$ cm$^{-2}$ s$^{-1}$
 (TS=26); where $F(>0.1$ GeV$)$ is the flux above 0.1 GeV. {\bf These
 values do not deviate by more than 2 sigmas from the P1234 average, 
$F (>0.1$ GeV$) = (13.8\pm1.8) \times 10^{-8}$ cm$^{-2}$ s$^{-1}$
(TS=100), and hence EGRET did not detect significant variability in
the flux from Mrk\,421.} We can easily
 obtain the \Fermi $F(>0.1$ GeV$)$ fluxes by using the flux (F) index ($\Gamma$) values
 reported in Figure~\ref{fig:Lc7days} (E$>$0.3 GeV): $F(>0.1$ GeV$) =
 F(>0.3$ GeV$)
 \times \left(0.3/0.1 \right)^{\Gamma-1}$. Applying this simple
 formalism one gets, for the max/min fluxes from Figure
 \ref{fig:Lc7days}, $F^{Max}(>0.1$ GeV$) = (25.7\pm4.7) \times 10^{-8}$
 cm$^{-2}$ s$^{-1}$ and $F^{Min} (>0.1$ GeV$) = (5.6\pm2.4) \times 10^{-8}$
 cm$^{-2}$ s$^{-1}$.  {\bf The maximum flux measured by EGRET
  and LAT are similar, although the {\it minimum} fluxes are not.
The LAT's larger effective area
compared to EGRET  permits detection of lower $\gamma$-ray fluxes.  
In any case, the EGRET
 and LAT fluxes are
comparable, which may indicate that Mrk\,421 is not as variable in the
MeV/GeV range as at other wavelengths, particularly X-rays and TeV
$\gamma$-rays \citep[e.g.][]{Wagner2008}.}

\begin{figure}
  \centering
 \includegraphics[width=6.0 in]{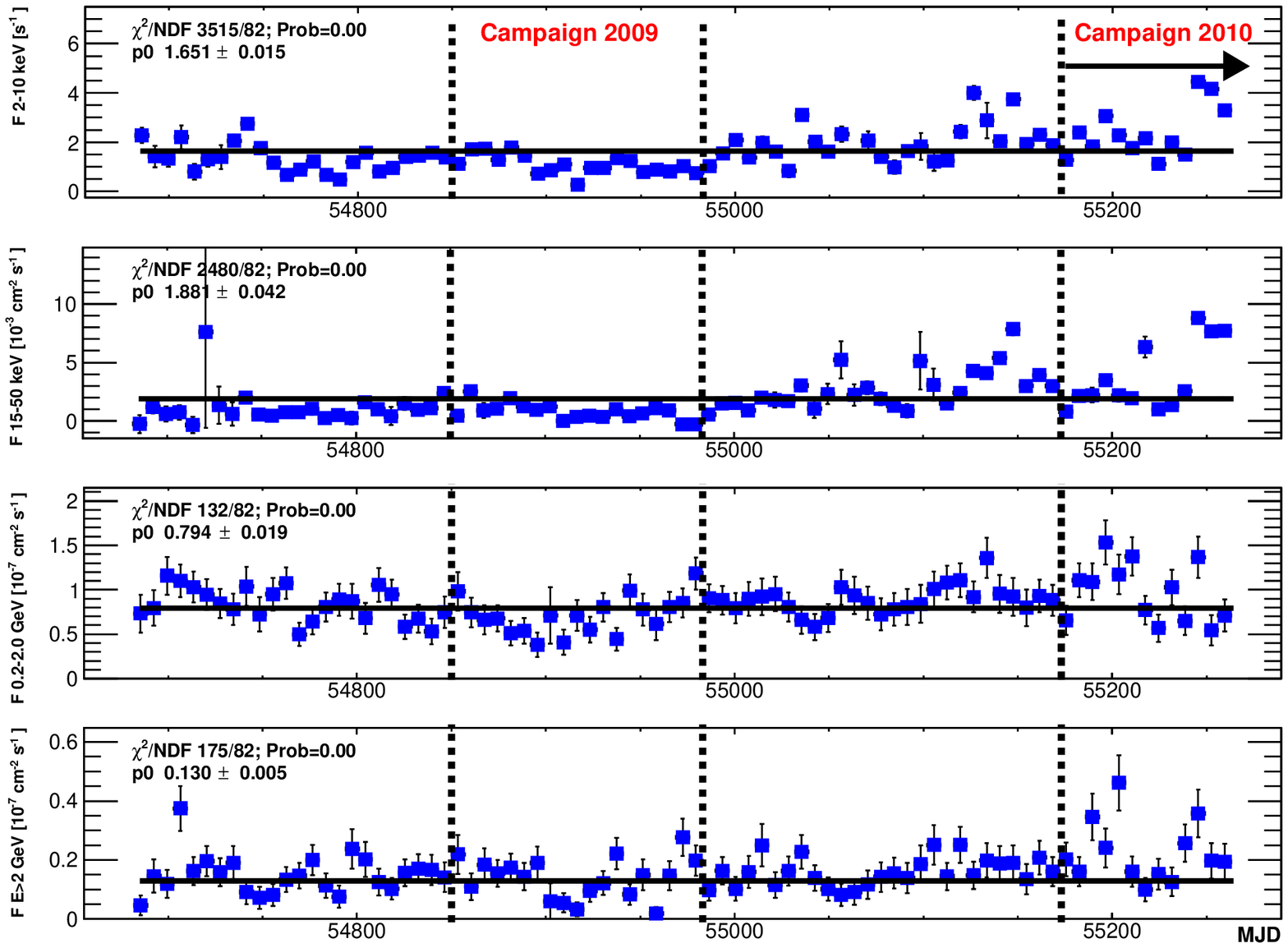}
  \caption{
Multifrequency light curves of Mrk\,421 with 7-day-long time bins obtained
with 3 all-sky-monitoring instruments: \RXTEc/ASM ($2-10$\,keV, first);
\Swiftc/BAT ($15-50$\,keV, second) and \FermiLAT for two different
energy ranges ($0.2-2$\,GeV, third, and $>2$\,GeV, fourth). The light
curves cover the period from 2008 August 5  (MJD 54683) to 2009 March 12 (MJD 55248). Vertical
bars denote $1\sigma$ uncertainties and horizontal error bars show the
width of the time interval. The black dashed lines and legends show
the results from constant fits to the entire data set. {\bf The
vertical dashed lines denote the time intervals with the extensive
multifrequency campaigns during the  2009 and 2010 seasons. }
}
 \label{fig:Lc7daysMW}
 \end{figure}

 The \FermiLAT capability for constant source monitoring is nicely
 complemented at X-ray energies by \RXTEc/ASM and \Swiftc/BAT, the two
 other all-sky instruments which can probe the X-ray activity of
 Mrk\,421 on 7-day-long time intervals. Figure~\ref{fig:Lc7daysMW} shows the
 measured fluxes by ASM in the energy range $2-10$\,keV, by BAT at
 $15-50$\,keV, and by LAT in two different energy bands: $0.2-2$\,GeV
 (low energy) and $>2$\,GeV (high energy)\footnote{The fluxes depicted
   in the light curves were computed fixing the photon index to 1.78
   (average index during the first 1.5 years of \Fermi operation) and
   fitting only the normalization factor of the power law function.}.
{\bf  
The low/high \FermiLAT energy bands were chosen (among other reasons) to
produce comparable flux errors. This might seem surprising at
first glance, given that the number of detected photons in the low
energy band is about 5 times larger than in the high energy band  
(for a differential energy spectrum parameterized by a
power law with photon index of 1.8, which is the case of
Mrk\,421). Hence the number of detected $\gamma$-rays decreases from about 50
down to about 10 for time intervals of 7
days. The main reason for having comparable flux errors in these two
energy bands is that the diffuse background, which follows a power law
with index 2.4 for the high galactic latitude of Mrk\,421, is about 25
times smaller in the high energy band.  Consequently, 
$Signal/Noise \sim N_S/\sqrt(N_B)$ remains approximately equal.}

{\bf  We do not see  variations in the LAT hardness
ratio (i.e. F($>2$\,GeV)/F($0.2-2$\,GeV) with the $\gamma$-ray flux, but
this is limited by the
relatively large  uncertainties and the low $\gamma$-ray flux variability
during this time interval.}
 The data from \RXTEc/ASM were obtained from the ASM web
 page\footnote{See \texttt{http://xte.mit.edu/ASM\us lc.html}}. We
 filtered out the data according to the provided prescription in the
 ASM web page, and
 made a weighted average of all the dwells (scan/rotation of the
 ASM Scanning Shadow Cameras lasting 90 seconds)
  from the 7-day-long time intervals defined for the \Fermi data. The data
 from \Swiftc/BAT were gathered from the BAT web page\footnote{See
   \texttt{http://swift.gsfc.nasa.gov/docs/swift/results/transients/}}.
 We retrieved the daily averaged BAT values and produced weighted
 average  for all the 7-day-long time intervals defined
 for the \Fermi data.

{\bf The X-ray flux from Mrk~421 was $\sim 1.7$~ct~s$^{-1}$  in ASM and
 $\sim 1.9 \times 10^{-3}$ ~ct~s$^{-1}$~cm$^{-2}$ in BAT. These fluxes correspond
 to $\sim$22 mCrab in ASM (1 Crab = 75 ct~s$^{-1}$) and 9 mCrab in BAT
 (1 Crab = 0.22 ct~s$^{-1}$~cm$^{-2}$), although given the recent
 reports on flux variability from the Crab Nebula
 \citep[see][]{Wilson2011,Abdo2011,Tavani2011}, the flux from the Crab
 Nebula is not a good absolute standard candle
 any longer and hence those numbers need to be taken with caveats.} One may note that the
 X-ray activity was rather low during the first year of Fermi
 operation. The X-ray activity increased around MJD 54990 and then 
 increased even more around MJD 55110. The $\gamma$-ray activity
 seemed to follow some of the X-ray activity, but the variations in
 the $\gamma$-ray range are substantially smoother than those observed
 in X-rays.

\begin{figure}
  \centering
  \includegraphics[width=6.0 in]{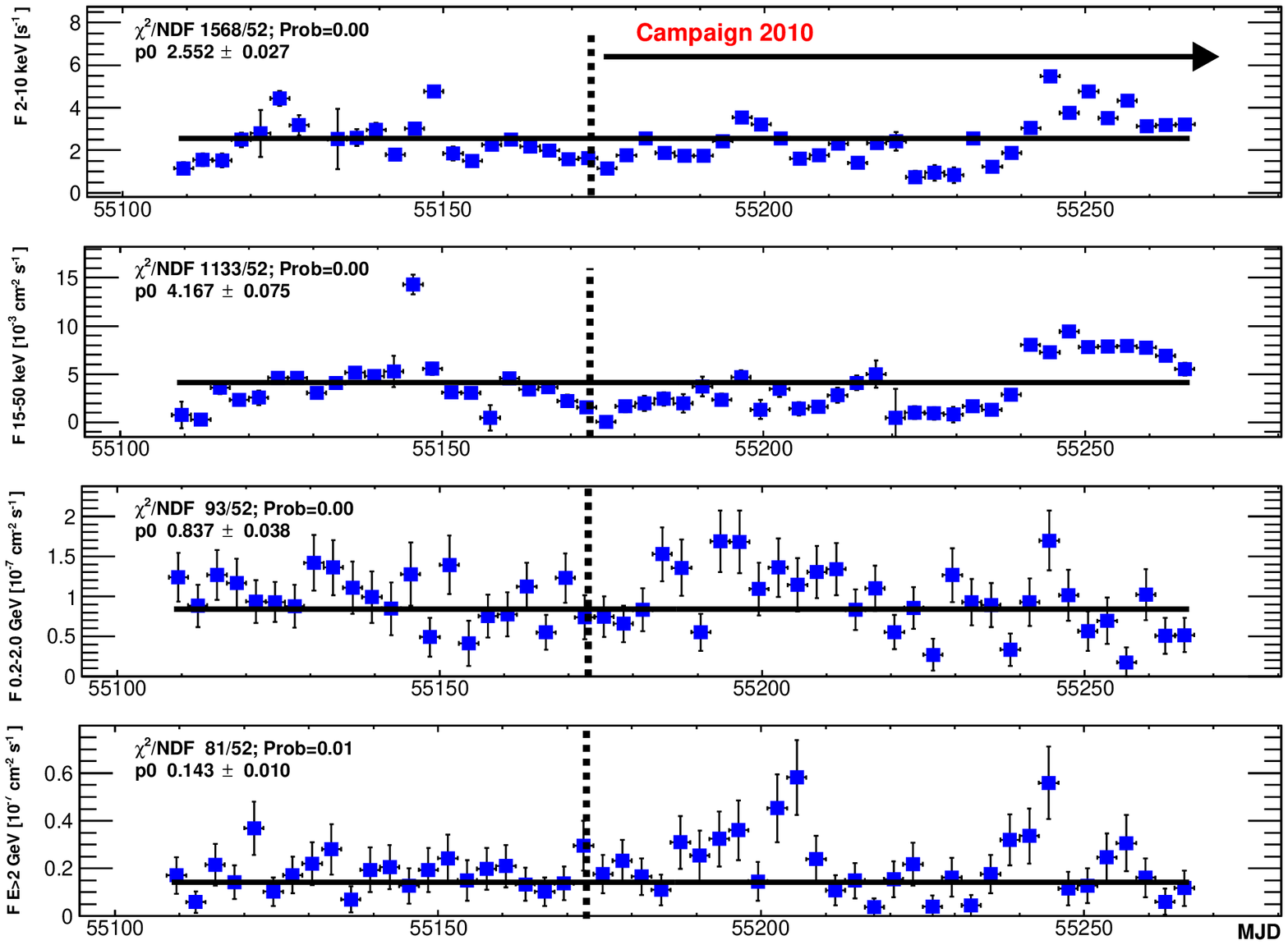}
  \caption{Multifrequency light curves of Mrk\,421 with 3-day-long
    time bins obtained with 3 all-sky-monitoring instruments:
    \RXTEc/ASM ($2-10$\,keV, first); \Swiftc/BAT ($15-50$\,keV,
    second) and \FermiLAT for two different energy ranges
    ($0.2-2$\,GeV, third, and $>2$\,GeV, fourth). The light curves
    cover the period from 2009 October 4 to 2010 March 12. Vertical bars denote $1\sigma$ uncertainties and horizontal error bars show the width of the time interval. The black dashed lines and legends show the results from constant fits to the entire data set. {\bf The
vertical dashed lines denote the beginning  of the extensive
multifrequency campaign on Mrk\,421 during the 2010 season. }}
 \label{fig:LcFlare_3day}
\end{figure}

Figure~\ref{fig:LcFlare_3day} shows the same light curves as Figure
\ref{fig:Lc7daysMW}, but only during the period of time after MJD
55110 (when Mrk\,421 showed high X-ray activity) with a time bin of
only 3 days. During this time period the ASM and BAT flux (integrated
over 3 days) went beyond 5~ct~s$^{-1}$ and $8 \times 10^{-3}$~ct~s$^{-1}$~cm$^{-2}$,
respectively, which implies a flux increase by a factor of 5-8 with
respect to the average fluxes during the first year. It is worth
noting that these large flux variations do not have a counterpart at
$\gamma$-ray energies measured by \FermiLAT. The MeV/GeV flux measured
by LAT remained roughly constant, with the exception of a flux
increase by a factor of $\sim$2 for the time interval around MJD
55180-55210 and around MJD 55240-55250, which was also seen by
\RXTEc/ASM and (to some extent) by \Swiftc/BAT.

We quantified the correlation among the light curves shown in Figure
\ref{fig:Lc7daysMW} and Figure \ref{fig:LcFlare_3day} following the
prescription from \citet{Edelson1988}. {\bf The results are shown in table
\ref{TableWithDCF} for a time lag of zero, which is the one giving the
largest DCF values. There is no indication of correlated activity at positive/negative time
lags in the DCF vs time plot for any of the used X-ray/$\gamma$-ray bands.
The advantage of using the DCF instead of the Pearson's correlation
coefficient is that the later does not consider the error in the
individual flux points, while the former does. In this particular
situation it is relevant to consider these errors because they are
sometimes comparable to the magnitude of the measured
flux variations. The main result is a
clear ($DCF/DCF_{error}\sim 4$) correlation between ASM and BAT, while
there is no indication of X-ray/$\gamma$-ray correlation ($DCF/DCF_{error} \lapp
2$). The correlation between the \FermiLAT fluxes below and
above 2 GeV is not significant ($DCF= 0.31 \pm 0.14$), which is
probably due to the low variability at $\gamma$-rays, together with
the relatively large flux errors for the individual 7-day-long and
3-day-long time intervals.}

\begin{deluxetable}{lcccccc}
\tabletypesize{\scriptsize}
\tablecolumns{7} 
\tablewidth{0pc}
\tablecaption{Discrete Correlation Function (DCF) computed using the
  flux values reported in Figure~\ref{fig:Lc7daysMW} (7-day-long time
  intervals, first 1.5 years of \Fermi operation) and
  Figure~\ref{fig:LcFlare_3day} (3-day-long time interval during the
  last 5 months, where the X-ray activity was high). The DCF values
  are given for time lag zero. }
\tablehead{ 
\colhead{Interval}        &\colhead{ASM-BAT}  &\colhead{ASM-$LAT_{<2\,GeV} $} &\colhead{ASM-$LAT_{>2\,GeV} $} &\colhead{BAT-$LAT_{<2\,GeV} $} & \colhead{BAT-$LAT_{>2\,GeV} $ }  & \colhead{$LAT_{<2\,GeV} - LAT_{>2\,GeV} $}
}  
\startdata 
7-day-long             &  $0.73 \pm 0.20$ &  $0.28 \pm 0.15$&  $0.35 \pm 0.14$&  $0.20 \pm 0.13$ &  $0.26 \pm 0.13$ &  $0.31 \pm 0.14$    \\ 
3-day-long             &  $0.65 \pm 0.13$ &  $0.01 \pm 0.18$&  $0.15 \pm 0.19$&  $-0.03 \pm 0.13$ &  $0.01 \pm 0.13$ &  $0.29 \pm 0.17$   \\ 
\enddata
\tablecomments{The DCF was computed as prescribed in \citet{Edelson1988}.}
\label{TableWithDCF}
\end{deluxetable}

We followed the prescription given in \cite{Vaughan2003} to quantify the flux variability by means of the fractional variability parameter $F_{var}$, as a function of energy. In order to account for the individual flux measurement errors ($\sigma_{err,i}$), we used the ``excess variance'' \citep{Nandra1997, Edelson2002} as an estimator of the intrinsic source variance. This is the variance after subtracting the expected contribution from measurement errors. For a given energy range, the $F_{var}$\ is calculated as
\begin{displaymath}
F_{var} = \sqrt{\frac{S^2 - <\sigma_{err}^2 >}{<F>^2}}
\end{displaymath}
where  $<F>$ is the mean photon flux, $S$ the standard deviation of the $N$ flux points, and 
$<\sigma_{err}^2>$ the average mean square error, all determined for a given energy bin.

\begin{figure}[t]
  \centering
 \includegraphics[width=4.0in]{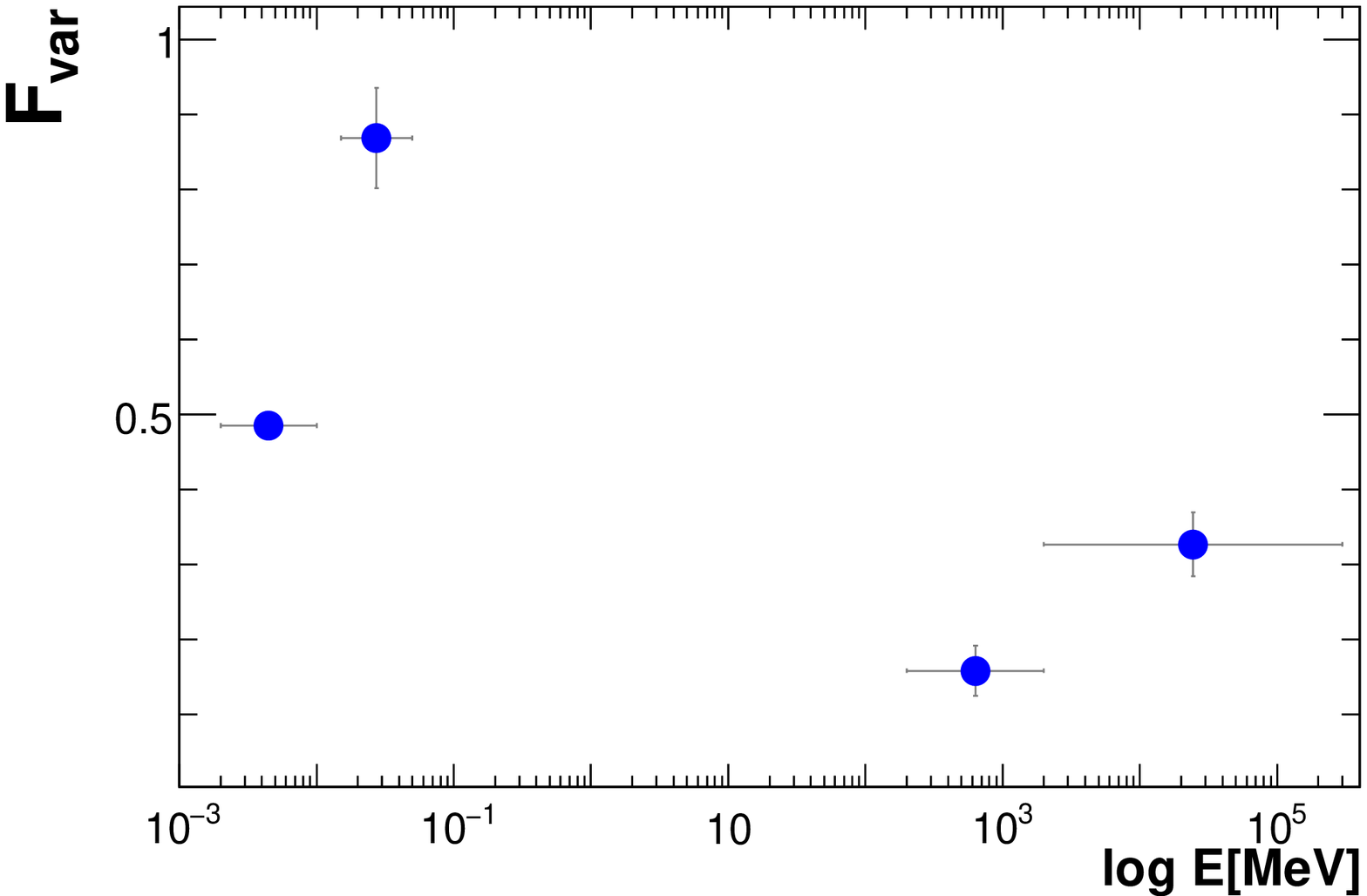}
   \caption{Fractional variability parameter for 1.5 year data (2008 August 5- 2009 March 12) from 3 all-sky-monitoring instruments: \RXTEc/ASM (2-10~keV, first); \Swiftc/BAT (15-50~keV,  second) and \FermiLAT  for 2 energy ranges 0.2-2 GeV and 2-300 GeV.  The fractional variability was computed according to \cite{Vaughan2003} using the light curves from Figure~\ref{fig:Lc7daysMW}. Vertical bars denote 1$\sigma$ uncertainties and horizontal bars indicate the width of each energy bin.}
 \label{fig:nva}
 \end{figure}

Figure~\ref{fig:nva} shows the derived $F_{var}$ values for the 4
energy ranges and time window covered by the light curves shown in
Figure~\ref{fig:Lc7daysMW}. The fractional variability is significant
for all energy ranges, with the X-rays having a substantially higher
variability than the $\gamma$-rays.  

It is interesting to note that, while the PL photon index variations
from Figure~\ref{fig:Lc7days} were not statistically significant
($\chi^{2}/NDF = 87/82$), Figure\ \ref{fig:nva} shows that the
fractional variability for photon energies above $2$\,GeV is higher
than that below $2$\,GeV; specifically $F_{var} (E<2$~GeV$) = 0.16 \pm
0.04$ while $F_{var} (E>2$GeV$) = 0.33 \pm 0.04$.  This apparent
discrepancy between the results reported in Figure\ \ref{fig:Lc7days}
and the ones reported in Figure \ref{fig:nva} (produced with the flux
points from Figure~\ref{fig:Lc7daysMW}) might be due to the fact that,
on time scales of 7-days, the photons below $2$\,GeV dominate the
determination of the PL photon index in the unbinned likelihood
fit. In other words, the source is bright enough in the energy range
$0.3-2$\,GeV such that the (relatively few) photons above $2$\,GeV do
not have a large (statistical) weight in the computation of the PL
photon index. Consequently, we are more sensitive to spectral
variations when doing the analysis separately for these two energy
ranges.

One may also note that, beside the larger variability in the \Fermi
fluxes above 2\,GeV with respect to those below 2\,GeV, the
variability in the BAT fluxes (15-50 keV) is also higher than that of
ASM (2-10 keV). The implications of this experimental result will be
further discussed in section \ref{Mrk421Variability}, in light of
the modeling results presented in section \ref{leptonicmodel}.

%% ============================================================================
%%
%% SECTION 4 -- Spectrum Analysis up to 400 GeV
%%
%% ============================================================================

%\vspace{-0.25cm}
\section{Spectral Analysis up to 400 GeV} 
\label{FermiSpectrum}

\begin{figure}
 \centering
\includegraphics[width=6.0 in]{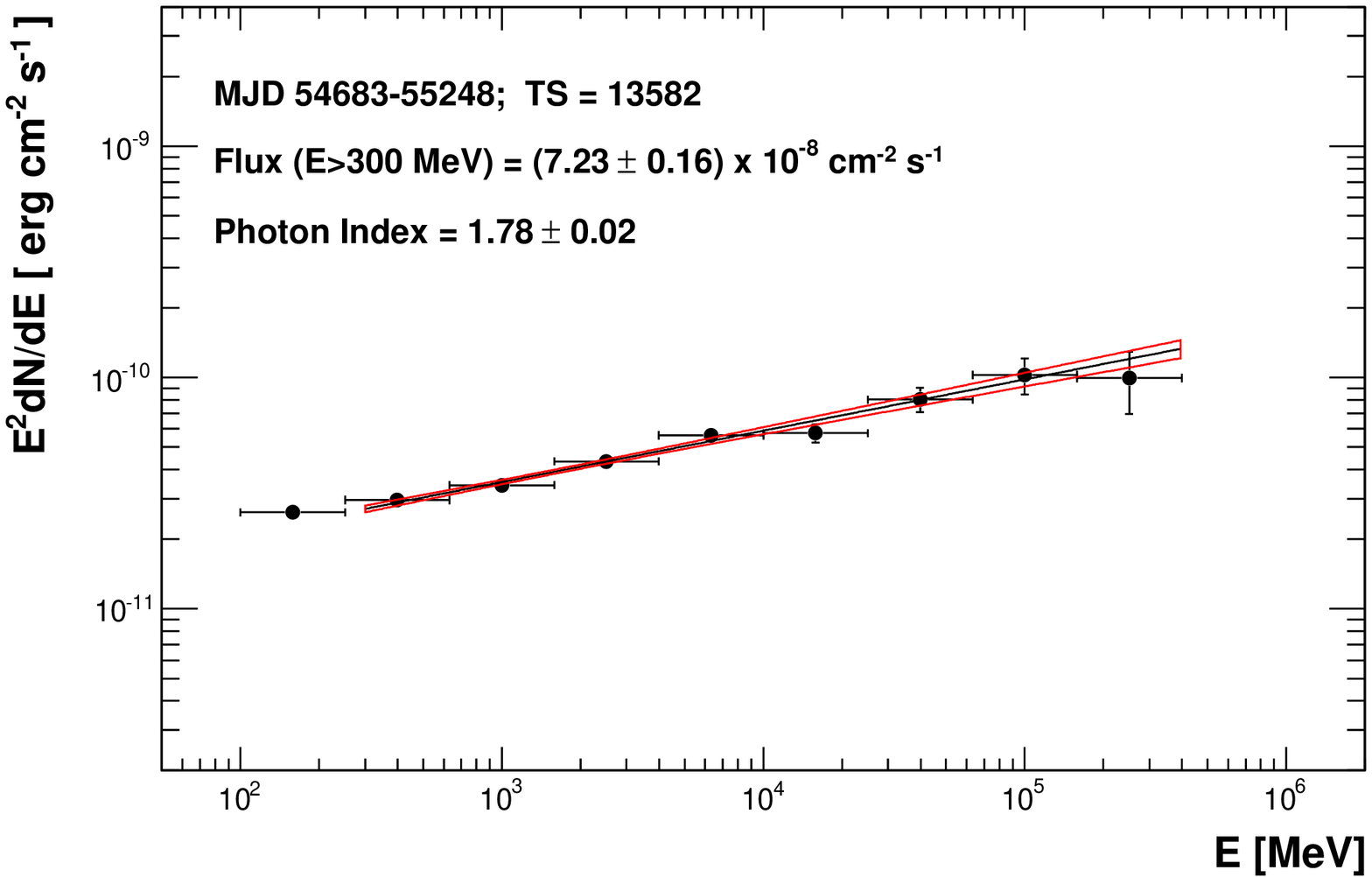}
  \caption{Fermi spectrum of Mrk\,421 during the period from 2008
    August 5 to 2010 February 20. Black line is the likelihood PL fit; red contour is 
the 68\% uncertainty of the fit and the black data points show the
energy fluxes computed on differential energy ranges. The inlay
summarizes the unbinned likelihood PL fit in the energy range 0.3-400 GeV. }
  \label{fig:SED}
 \end{figure}

The LAT instrument allows one to accurately reconstruct the photon
energy over many orders of magnitude. Figure~\ref{fig:SED} shows the
spectrum of Mrk\,421 in the energy range 0.1-400 GeV.  This is the
first time that the spectrum of Mrk\,421 can be studied with this
level of detail over this large fraction of the electromagnetic
spectrum, which includes the previously unexplored energy range 10-100
GeV. The spectrum was computed using the analysis procedures described
in section \ref{FermiData}. In order to reduced systematics, the
spectral fit was performed using photon energies greater than $0.3$
GeV, where the LAT instrument has good angular resolution and large
effective area. The black line in Figure~\ref{fig:SED} is the result
of a fit with a single PL function over the energy range 0.3-400 GeV,
and the red contour is the 68\% uncertainty of the fit. The data are
consistent with a pure PL function with photon index of $1.78 \pm
0.02$.  The black data points are the result of performing the
analysis on differential energy ranges (2.5 bins per decade of
energy)\footnote{Because of the analysis being carried out in small
energy ranges, we fixed the spectral index to $1.78$, which
is the value obtained when fitting the entire energy range. We
repeated the same procedure fixing the photon indices to 1.5 and 2.0,
and found no significant change. Therefore, the results from the
differential energy analysis are not sensitive to the selected photon
index used in the analysis.}.  The points are well within 1-2$\sigma$
from the fit to the overall spectrum (black line), which confirms that
the entire \Fermi spectrum is consistent with a pure power law
function.

However, it is worth noticing that the error bars at the highest
energies are relatively large due to the low photon count.  In the
energy bins 60-160 GeV and 160-400 GeV, the predicted (by the model
for Mrk\,421) number of photons detected by LAT is 33 and 11,
respectively.  Even though the low background makes those signals very
significant (TS values of 562 and 195 respectively), the statistical
uncertainties in the energy flux values are naturally large and hence
they could hide a potential turnover in the spectrum of Mrk\,421 at
around 100 GeV. Indeed, when performing the likelihood analysis on LAT
data above 100 GeV, one obtains a photon flux above 100 GeV of $(5.6
\pm 1.1) \times 10^{-10}$\,ph\,cm$^{-2}$\, s$^{-1}$ with a photon
index of $2.6 \pm 0.6$, which might suggest a turnover in the
spectrum, consistent with the TeV spectra determined by past
observations with IACTs
\citep{Krennrich2002,Aharonian2003,Aharonian2005,Mrk421MAGIC}.  In
order to make a statistical evaluation of this possibility, the LAT
spectrum (in the range 0.3-400 GeV) was fit with a broken power law
(BPL) function, obtaining the indices of $1.77 \pm 0.02$ and $2.9 \pm
1.0$ below and above the break energy of $182 \pm 39$ GeV,
respectively. The likelihood ratio of the BPL and the PL gave 0.7,
which, given the 2 extra degrees of freedom for the BPL function,
indicates that the BPL function is not statistically preferred over
the PL function.  Therefore, the statistical significance of the LAT
data above 100 GeV is not sufficiently high to evaluate the potential
existence of a break (peak) in the spectrum.

%% ============================================================================
%%
%% SECTION 5 -- Spectral Energy Distribution of Mrk~421 during the 4.5 month long multifrequency campaign from 2009
%%
%% ============================================================================

\section{Spectral Energy Distribution of Mrk\,421 during the 4.5 month long Multifrequency Campaign from 2009} 
\label{MWSED}

As mentioned in \S\ref{Intro}, we organized a multifrequency
(from radio to TeV photon energies) campaign to monitor Mrk\,421
during a time period of 4.5 months. The observing campaign started on
2009 January 19 (MJD 54850) and finished on 2009 June 1 (MJD
54983). The observing strategy for this campaign was to sample the
broad-band emission of Mrk\,421 every 2 days, which was accomplished
at optical, X-ray and TeV energies when the weather and/or technical limitations allowed. The main goal of this project
was to collect an extensive multifrequency data set that is
simultaneous and representative of the average/typical SED from
Mrk\,421. Such a data set can provide additional constraints that will
allow us to refine the emission models, which in turn will provide new
insights into the processes related to the particle acceleration and
radiation in this source.
In this section we describe the source coverage
during the campaign, the data analysis for several of the
participating instruments, and finally we report on the averaged SED
resulting from the whole campaign. 

\subsection{Details of the Campaign: Participating Instruments  and Temporal Coverage}

\begin{figure}[t]
  \centering
 \includegraphics[width=6.5in]{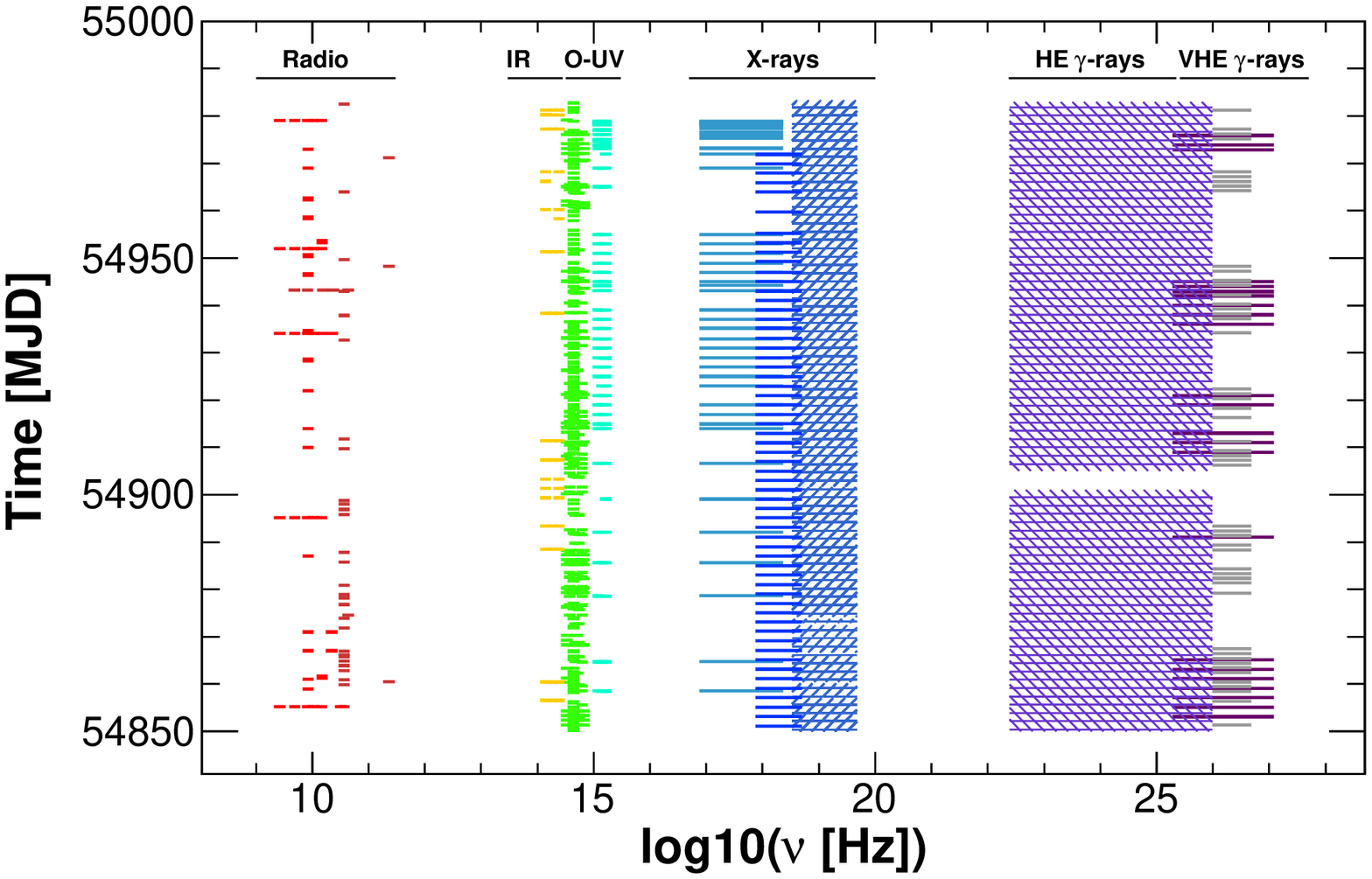}
   \caption{Time and Energy coverage during the multifrequency
     campaign. For the sake of clarity, the minimum observing time
     displayed in the plot was set to half a day. }
  \label{fig:TimeEnergyCoverage}
\end{figure}

The list of all the instruments that participated in the campaign are
reported in Table\,\ref{TableWithInstruments}, and the scheduled
observations can be found online\footnote{See
\texttt{https://confluence.slac.stanford.edu/display/GLAMCOG/Campaign+on+Mrk421+(Jan+2009+to+May+2009))}
maintained by D. Paneque.}. We note that in some cases the planned
observations could not be performed due to bad observing conditions,
while on other occasions the observations were performed but the data
could not be properly analyzed due to technical problems or rapidly
changing weather conditions.  Figure\,\ref{fig:TimeEnergyCoverage}
shows the time coverage as a function of the energy range for the
instruments/observations used to produce the SED shown in
Figure\,\ref{fig:MWSED}. Apart from the unprecedented energy coverage
(including, for the first time, the GeV energy range from \FermiLATc),
the source was sampled very uniformly with the various instruments
participating in the campaign and, consequently, it is reasonable to
consider the SED constructed below as the actual average (typical) SED
of Mrk\,421 during the time interval covered by this multifrequency
campaign. The largest non-uniformity in the sampling of the source
comes from the Cherenkov Telescopes, which are the instruments most
sensitive to weather conditions. Moreover, while there are many
radio/optical instruments spread all over the globe, in this observing
campaign, only two Cherenkov Telescope observatories participated,
namely MAGIC and Whipple. Hence, the impact of observing conditions
was more important to the coverage at the VHE $\gamma$-ray
energies. During the time interval MJD54901-54905 the \Fermi satellite
did not operate due to a spacecraft technical problem.  The lack of
\FermiLAT data during this period is clearly seen in
Figure\,\ref{fig:TimeEnergyCoverage}.

We note that Figure\,\ref{fig:TimeEnergyCoverage} shows the MAGIC and
Whipple coverage in VHE $\gamma$-ray energies, but only the MAGIC
observations were used to produce the spectra shown in
Figure\,\ref{fig:MWSED}.  The more extensive, but less sensitive,
Whipple data (shown as grey boxes in
Figure\,\ref{fig:TimeEnergyCoverage}) were primarily taken to
determine the light curve \citep{Pichel2009} and a re-optimization was
required to derive the spectrum, which will be reported elsewhere.

In the following paragraphs we briefly discuss the procedures used in the analysis of the instruments participating in the campaign. The analysis of the \FermiLAT data was described in \S\ref{FermiData} and the results obtained will be described in detail in \S\ref{FermiSED_InMW}.

\subsubsection{Radio Instruments}
\label{radio}

Radio data were taken for this campaign from single-dish telescopes,
one mm-interferometer, and one Very Long Baseline Interferometry
(VLBI) array, at frequencies between $2.6$\,GHz and $225$\,GHz (see
Table\,\ref{TableWithInstruments}). The single-dish telescopes were
the Effelsberg 100\,m radio telescope, the 32\,m Medicina radio
telescope, the 14\,m Mets\"ahovi radio telescope, the 32\,m Noto radio
telescope, the Owens Valley Radio Observatory (OVRO) 40\,m telescope,
the 26\,m University of Michigan Radio Astronomy Observatory
(UMRAO) and the 600 meter ring radio telescope RATAN-600.  
The mm-interferometer was the Sub-millimeter Array (SMA). The
NRAO Very Long Baseline Array (VLBA) was used for the VLBI
observations. For the single-dish instruments and SMA, Mrk\,421 is
pointlike and unresolved at all observing frequencies. Consequently,
the single-dish measurements denote the total flux density of the
source integrated over the whole source extension. Details of the
observing strategy and data reduction are given by \citet[F-GAMMA
project]{Fuhrmann2008,Angelakis2008},
\citet[Mets\"ahovi]{Terasranta1998}, \citet[UMRAO]{Aller1985},
\citet[Medicina and Noto]{Venturi2001},  \citet[RATAN-600]{Kovalev1999} and \citet[in preparation, OVRO]{Richards2010}.

The VLBA data were obtained at various frequencies (5, 8, 15, 24,
43\,GHz) through various programs (BP143, BK150 and MOJAVE). The data
were reduced following standard procedures for data reduction and
calibration \citep[see, for example,][for a description of the MOJAVE
and BK150 programs, respectively]{Lister2009,Sokolovsky2010}. Since
the VLBA angular resolution is smaller than the radio source
extension, measurements were performed for the most compact core
region, as well as for the total radio structure at parsec scales. The
core is partially-resolved by our 15, 24 and 43 GHz observations
according to the resolution criterion proposed by \citet{Kovalev2005}
and \citet{Lobanov2005}.  The VLBA core size was determined with
two-dimensional Gaussian fits to the measured visibilities. The FWHM
size of the core was estimated to be in the range 0.06--0.12 mas at
the highest observing frequencies, 15, 24 and 43\,GHz.  Both the total
and the core radio flux densities from the VLBA data are shown in
Figure\,\ref{fig:MWSED}.

\subsubsection{Optical and Near-IR Instruments}

The coverage at optical frequencies was provided  by various
telescopes around the globe, and this decreased the sensitivity to
weather and technical difficulties and provided good overall coverage
of the source, as depicted in
Figure\,\ref{fig:TimeEnergyCoverage}. Many of the observations were
performed within the GASP-WEBT program \citep[e.g.,][]{Villata2008,
Villata2009}; this is the case for the data collected by the
telescopes at Abastumani, Lulin, Roque de los Muchachos (KVA),
St. Petersburg, Talmassons, and Valle d'Aosta observatories ($R$
band). In addition, the telescopes GRT, ROVOR, New Mexico Skies and
MitSume provided data with various optical filters, while OAGH and
WIRO provided data at near-infrared wavelengths. See
Table\,\ref{TableWithInstruments} for further details.

All the optical and near/IR instruments used the calibration stars
reported in \citet{Villata1998}, and the Galactic extinction was
corrected with the coefficients given in \citet{schlegel98}. The flux
from the host galaxy (which is significant only below $\nu \sim
10^{15}~Hz$) was estimated using the flux values at the $R$ band from
\cite{Nilsson2007} and the colors reported in \cite{Fukugita1995}, and
then subtracted from the measured flux.

\subsubsection{\Swiftc/UVOT}

The \Swift Ultraviolet/Optical Telescope \citep[UVOT;][]{Roming2005}
dataset includes all the observations performed during the
time interval MJD 54858 to 54979, which amounts to 46 single pointing
observations that were requested to provide UV coverage during the
Mrk\,421 multifrequency campaign. The UVOT telescope cycled through
each of three ultraviolet passbands (UVW1, UVM2, UVW2). Photometry was
computed using a $5$\,arcsec source region around Mrk\,421 using a
custom UVOT pipeline that performs the calibrations presented in
\citet{Poole2008}. Moreover, the custom pipeline also allows for
separate, observation-by-observation, corrections for astrometric
mis-alignments \citep[][in preparation]{AcciariMrk4212008}.  A visual
inspection was also performed on each of the observations to ensure
proper data quality selection and correction.  The flux measurements
obtained have been corrected for Galactic extinction $E_{B-V} =
0.019$\,mag \citep[]{schlegel98} in each spectral band
\citep[]{Fitzpatrick99}.

\subsubsection{\Swiftc/XRT} 

All the \Swift X-ray Telescope \citep[XRT;][]{Burrows2005} Windowed
Timing observations of Mrk\,421 carried out from MJD 54858 to 54979
were used for the analysis: this amounts to a total of 46 observations
that were performed within this dedicated multi-instrument effort. The
XRT data set was first processed with the XRTDAS software package
(v.2.5.0) developed at the ASI Science Data Center (ASDC) and
distributed by HEASARC within the HEASoft package (v.6.7). Event files
were calibrated and cleaned with standard filtering criteria with the
{\em xrtpipeline} task using the latest calibration files available in
the \Swift CALDB. The individual XRT event files were then merged
together using the XSELECT package and the average spectrum was
extracted from the summed event file. Events for the spectral analysis
were selected within a circle with a 20 pixel ($\sim 47$\,arcsec) radius,
which encloses about $95\%$ of the PSF, centered on the source
position. The background was extracted from a nearby circular region
of 40 pixel radius. The source spectrum was binned to ensure a minimum
of 20 counts per bin to utilize the $\chi^{2}$ minimization fitting
technique.  {\bf In addition, we needed to apply a small energy
offset ($\sim 40$\,eV) to the observed energy spectrum. The origin of
this correction is likely to be CCD charge traps generated by
radiation and high-energy proton damage (SWIFT-XRT-CALDB-12), which
affect mostly the lowest energies (first 1-2 bins) of the
spectrum.} The ancillary response files were generated with the {\em
xrtmkarf} task applying corrections for the PSF losses and CCD defects
using the cumulative exposure map. The latest response matrices
(v.011) available in the \Swift CALDB were used.

The XRT average spectrum in the $0.3-10$\,keV energy band was fitted
using the XSPEC package. We adopted a log-parabolic model of the form
$F(E)=K \cdot (\frac{E}{keV})^{-(\Gamma + \beta \cdot log(\frac{E}{keV}))}$
\citep{Massaro2004a,Massaro2004b} with an absorption
hydrogen-equivalent column density fixed to the Galactic value in the
direction of the source, which is  $1.61 \times 10^{20}$\,cm$^{-2}$
\citep{Kalberla2005}. We found that this model provided a good
description of the observed spectrum, with the exception of the
$1.4-2.3$\,keV energy band where spectral fit residuals were
present. These residuals are due to known XRT calibration
uncertainties (SWIFT-XRT-CALDB-12\footnote{
\texttt{http://heasarc.gsfc.nasa.gov/docs/heasarc/caldb/swift/docs/xrt/SWIFT-XRT-CALDB-09\us
v12.pdf}}) and hence we decided to exclude the $1.4-2.3$\,keV energy
band from the analysis. The resulting spectral fit gave the following parameters: $K
= (1.839 \pm 0.002) \times 10^{-1}$ ph cm$^{-2}$ s$^{-1}$ keV$^{-1}$,
$\Gamma=2.178 \pm 0.002$, $\beta= 0.391 \pm 0.004$.  The XRT SED data shown in
figure \ref{fig:MWSED} were corrected for the Galactic absorption and
then binned in 16 energy intervals.

\subsubsection{\RXTEc/PCA}

The {\em Rossi} X-ray Timing Explorer \citep[\RXTEc;][]{RXTERef}
satellite performed 59 pointing observations of Mrk\,421 during the
time interval  MJD 54851 and 54972. These observations amount to a
total exposure of 118 ks, which was requested through a dedicated
Cycle 13 proposal to provide X-ray coverage for this multi-instrument
campaign on Mrk\,421.

The data analysis was performed using \texttt{FTOOLS} v6.9 and
following the procedures and filtering criteria recommended by the
\RXTE Guest Observer
Facility\footnote{\url{http://www.universe.nasa.gov/xrays/programs/rxte/pca/doc/bkg/bkg-2007-saa/}}
after September 2007.  {\bf The  average net count rate from Mrk\,421 was
about 25\,ct s$^{-1}$ per pcu (in the energy range $3-20$\,keV) 
with flux variations typically much smaller than a factor of two. }Consequently,
the observations were filtered
following the conservative procedures for faint sources: Earth elevation angle greater than $10^\circ$, pointing offset less than $0.02^{\circ}$, time since the peak of the last SAA (South Atlantic Anomaly) passage greater than 30 minutes, and electron contamination less than $0.1$. For further details on the analysis of faint sources with \RXTEc, see the online Cook Book\footnote{\url{http://heasarc.gsfc.nasa.gov/docs/xte/recipes/cook_book.html}}. In the data analysis, in order to increase the quality of the signal, only the first xenon layer of PCU2 was used. We used the package \texttt{pcabackest} to model the background and the package \texttt{saextrct} to produce spectra for the source and background files and the script\footnote{The CALDB files are located at \url{http://heasarc.gsfc.nasa.gov/FTP/caldb}}  \texttt{pcarsp} to produce the response matrix.

The PCA average spectrum in the $3-32$\,keV energy band was fitted
using the XSPEC package using a PL function with an exponential cutoff
(cutoffpl) with a non-variable neutral Hydrogen column density $N_H$
fixed to the Galactic value in the direction of the source ($1.61
\times 10^{20}$\,cm$^{-2}$; \citep{Kalberla2005}). However, since the
PCA bandpass starts at 3 keV, the value for $N_H$ used does not
significantly affect our results.  The resulting spectral fit provided
a good representation of the data for the following parameters:
normalization parameter $K = (2.77 \pm 0.03) \times 10^{-1}$ ph
cm$^{-2}$ s$^{-1}$ keV$^{-1}$, photon index $\Gamma = 2.413 \pm 0.015$, and
cutoff energy $E_{exp}=22.9 \pm 1.3$ keV.  The obtained 23 energy bins
PCA average spectrum is shown in Figure~\ref{fig:MWSED}.

\subsubsection{\Swiftc/BAT} 

The \Swift Burst Alert Telescope \citep[BAT;][]{Barthelmy2005}
analysis results presented in this paper were derived with all the
available data during the time interval MJD 54850 and 54983.  The
spectrum was extracted following the recipes presented in
\citet[][]{ajello08,ajello09b}. This spectrum is constructed by weight
averaging the source spectra extracted over short exposures
(e.g. 300\,s) and it is representative of the averaged source emission
over the time range spanned by the observations. These spectra are
accurate to the mCrab level and the reader is referred to
\cite{ajello09a} for more details. The \Swiftc/BAT spectrum in the
15--200 keV energy range is consistent with a PL function with
normalization parameter $K = 0.46 \pm 0.27$ ph~cm$^{-2}$ s$^{-1}$
keV$^{-1}$ and photon index $\Gamma = 3.0 \pm 0.3$. {\bf The last two
  flux points are within one standard deviation from the above
  mentioned PL function and hence the apparent 
upturn given by these last two data points in the spectrum is not significant.}

\subsubsection{MAGIC}

MAGIC is a system of two 17\,m-diameter IACTs for VHE $\gamma$-ray
astronomy located on the Canary Island of La Palma, at an altitude of
2200\,m above sea level. At the time of the observation, MAGIC-II, the
new second telescope of the current array system, was still in its
commissioning phase so that Mrk\,421 was observed in stand-alone mode
by MAGIC-I, which is in scientific operation since 2004
\citep{Albert2008}.  The MAGIC observations were performed in the
so-called ``wobble'' mode \citep{Daum1997}.  In order to have a low
energy threshold, only observations at zenith angles less than
$35^{\circ}$ were used in this analysis.  The bad weather and a shut
down for a scheduled hardware system upgrade during the period MJD
54948-54960 (April 27--May 13) significantly reduced the amount of
time that had initially been scheduled for this campaign. The data
were analyzed following the prescription given by \citet{Albert2008}
and \citet{Aliu2009}. The data surviving the quality cuts amounted to
a total of 27.7 hours. The preliminary reconstructed photon fluxes for
the individual observations gave an average flux of about 50\% that of
the Crab Nebula, with relatively mild (typically less than factor 2)
flux variations.  The derived spectrum was unfolded to correct for the
effects of the limited energy resolution of the detector and possible
bias \citep{Albert2007c}. The resulting spectrum was fit satisfactorily
with a single log-parabola function: $F(E)=K \cdot (\frac{E}{0.3 {\rm
TeV}})^{-(\Gamma +\beta \cdot log(\frac{E}{0.3 {\rm TeV}}))}$. The resulting
spectral fit gave the following parameters: $K = (6.50 \pm 0.13)
\times 10^{-10}$ ph~cm$^{-2}$ s$^{-1}$ erg$^{-1}$, $\Gamma=2.48 \pm 0.03$,
$\beta= 0.33 \pm 0.06$, with $\chi ^2/NDF = 11/6$. A fit with a simple
power-law function gives $\chi ^2/NDF=47/7$, which confirmed the
existence of curvature in the VHE spectrum.

\begin{deluxetable}{lll}
\rotate
\tabletypesize{\scriptsize}
\tablecolumns{3} 
\tablewidth{0pc}
\tablecaption{List of instruments participating in the multifrequency campaign and used in the compillation of the SED shown in Figure\,\ref{fig:MWSED}}
\tablehead{ 
\colhead{Instrument/Observatory}                   &\colhead{Energy range covered}  &\colhead{Web page} 
}  
\startdata 
MAGIC                 & 0.08-5.0\,TeV               & \url{http://wwwmagic.mppmu.mpg.de/} \\
Whipple$^{a}$              &0.4-2.0\,TeV                  & \url{http://veritas.sao.arizona.edu/content/blogsection/6/40/} \\           
\FermiLAT                & 0.1-400\,GeV               & \url{http://www-glast.stanford.edu/index.html} \\
\Swiftc/BAT                & 14-195\,keV               & \url{http://heasarc.gsfc.nasa.gov/docs/swift/swiftsc.html/} \\
\RXTEc/PCA                & 3-32\,keV               & \url{http://heasarc.gsfc.nasa.gov/docs/xte/rxte.html} \\
\Swiftc/XRT                & 0.3-9.6\,keV               & \url{http://heasarc.gsfc.nasa.gov/docs/swift/swiftsc.html} \\
\Swiftc/UVOT                & UVW1, UVM2, UVW2            & \url{http://heasarc.gsfc.nasa.gov/docs/swift/swiftsc.html} \\
Abastumani {\scriptsize (through GASP-WEBT program)}              & R       band        & \url{http://www.oato.inaf.it/blazars/webt/} \\
Lulin {\scriptsize (through GASP-WEBT program)}              & R       band        & \url{http://www.oato.inaf.it/blazars/webt/} \\
Roque de los Muchachos (KVA) {\scriptsize (through GASP-WEBT program)}              & R       band        & \url{http://www.oato.inaf.it/blazars/webt/} \\
St. Petersburg {\scriptsize (through GASP-WEBT program)}              & R       band        & \url{http://www.oato.inaf.it/blazars/webt/} \\
Talmassons {\scriptsize (through GASP-WEBT program)}              & R       band        & \url{http://www.oato.inaf.it/blazars/webt/} \\
Valle d'Aosta {\scriptsize (through GASP-WEBT program)}              & R       band        & \url{http://www.oato.inaf.it/blazars/webt/} \\
GRT                & V, R, B, I bands                & \url{http://asd.gsfc.nasa.gov/Takanori.Sakamoto/GRT/index.html} \\
ROVOR                &   B, R, V  bands            & \url{http://rovor.byu.edu/} \\
New Mexico Skies                &   R, V bands            & \url{http://www.nmskies.com/equipment.html/} \\
MitSume                & g, Rc, Ic bands               & \url{http://www.hp.phys.titech.ac.jp/mitsume/index.html} \\
OAGH    &  H, J, K bands             & \url{http://astro.inaoep.mx/en/observatories/oagh/} \\
WIRO                & J, K bands            & \url{http://physics.uwyo.edu/~chip/wiro/wiro.html} \\
SMA      &   225\,GHz  &  \url{http://sma1.sma.hawaii.edu/} \\
VLBA  &  4.8, 8.3, 15.4, 23.8, 43.2\,GHz               & \url{http://www.vlba.nrao.edu/} \\
Noto & 8.4, 22.3\,GHz & \url{http://www.noto.ira.inaf.it/} \\
Mets\"ahovi     {\scriptsize (through GASP-WEBT program)}                   & 37\,GHz               & \url{http://www.metsahovi.fi/} \\
VLBA  {\scriptsize (through MOJAVE program)}                  &  15\,GHz               & \url{http://www.physics.purdue.edu/MOJAVE/} \\
OVRO                & 15\,GHz              & \url{http://www.ovro.caltech.edu/} \\
Medicina & 8.4 \,GHz & \url{http://www.med.ira.inaf.it/index_EN.htm} \\
UMRAO {\scriptsize (through GASP-WEBT program)}                & 4.8, 8.0, 14.5\,GHz               & \url{http://www.oato.inaf.it/blazars/webt/} \\
RATAN-600  &  2.3, 4.8, 7.7, 11.1, 22.2 GHz & \url{http://w0.sao.ru/ratan/} \\
Effelsberg {\scriptsize (through F-GAMMA program)}               &  2.6, 4.6, 7.8, 10.3, 13.6, 21.7, 31\,GHz             & \url{http://www.mpifr-bonn.mpg.de/div/effelsberg/index_e.html/} \\

\enddata
\tablecomments{The energy range shown in column 2 is the actual energy range covered during the Mrk\,421 observations, and not the instrument nominal energy range, which might only be achievable for bright sources and excellent observing conditions. }
\tablecomments{$(a)$ The Whipple spectra were not included in Figure\,\ref{fig:MWSED}. See text for further comments.}
\label{TableWithInstruments}
\end{deluxetable}

\subsection{\FermiLAT Spectra During the Campaign}
\label{FermiSED_InMW}

\begin{figure}[t]
  \centering
  \includegraphics[height=2.2 in]{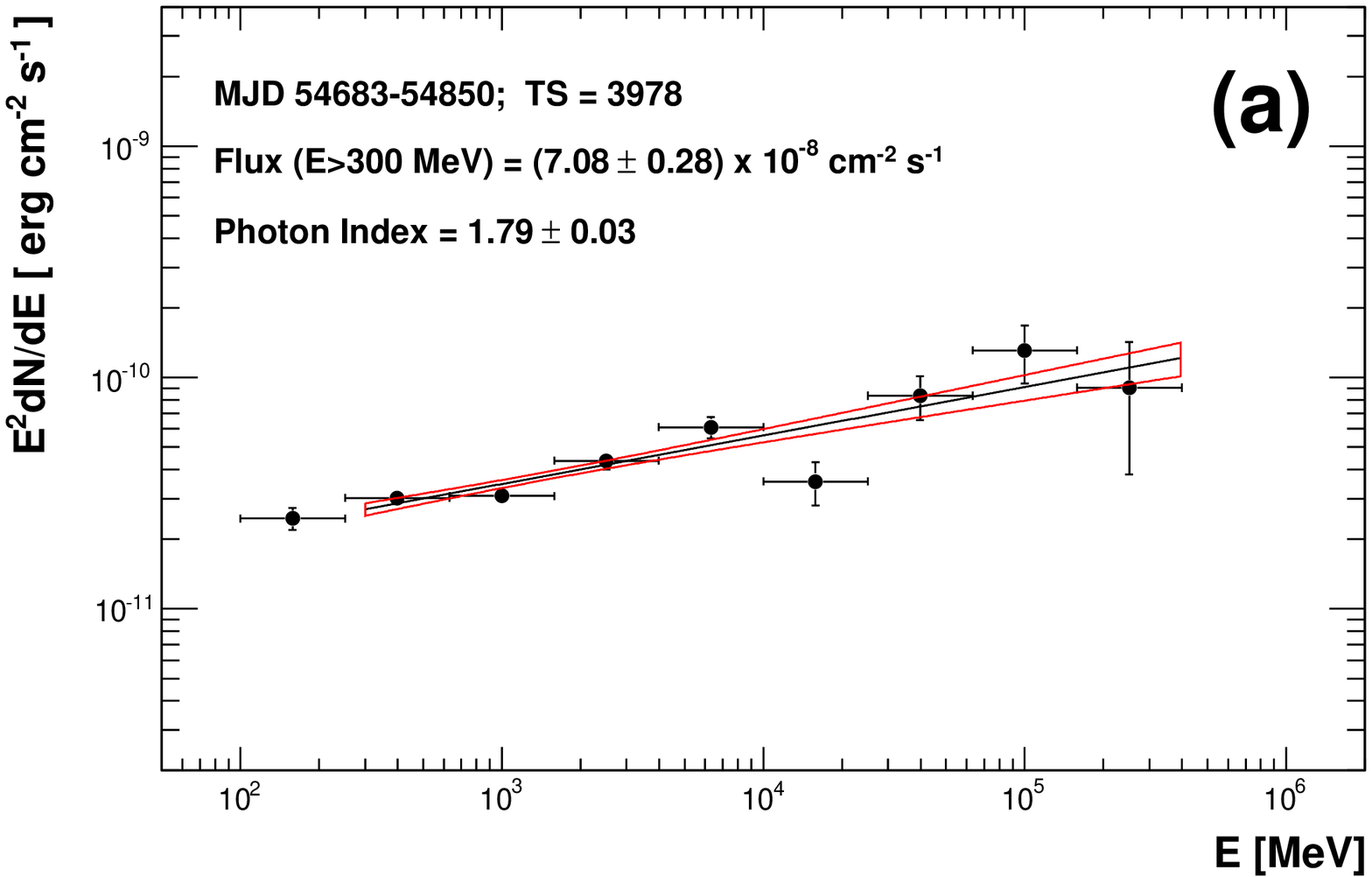}
  \includegraphics[height=2.2 in]{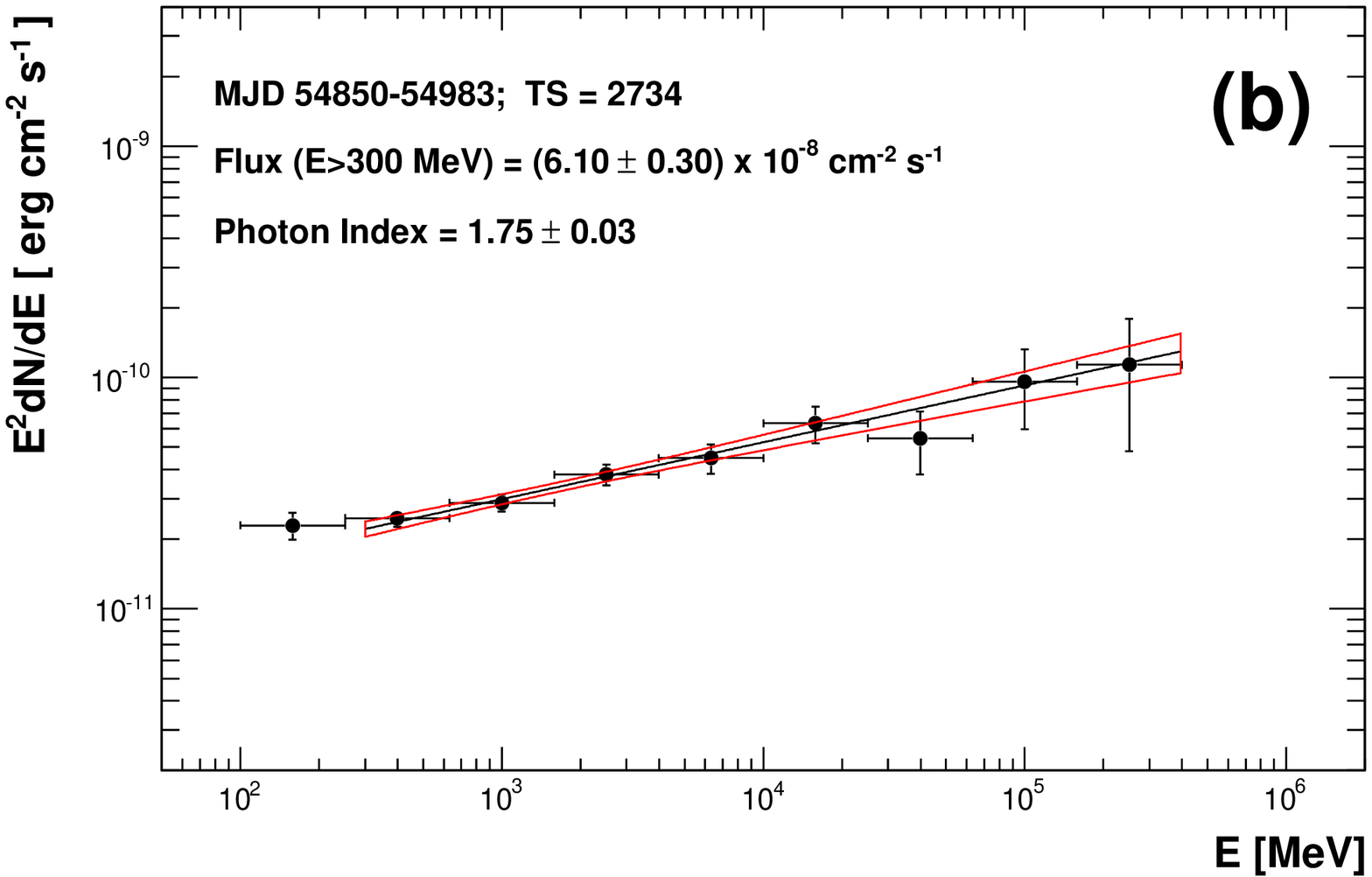}
  \includegraphics[height=2.2 in]{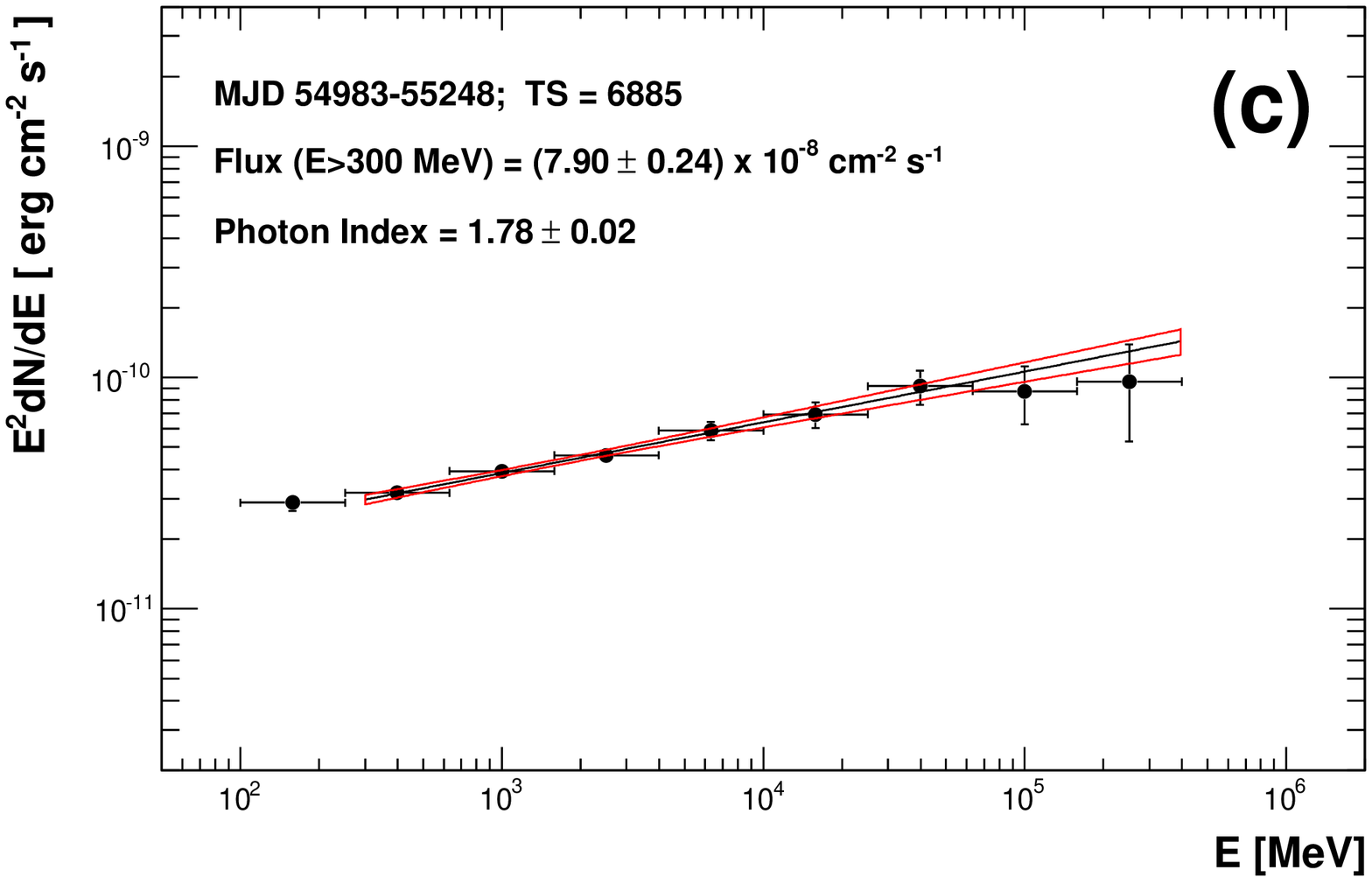}
  \caption{Fermi spectra of Mrk\,421 for several time
    intervals of interest. The panel {\em (a)} shows the spectrum for the time
    period before the multifrequency campaign (MJD 54683-54850), the
    panel {\em (b)} for the time interval corresponding to the
    multifrequency campaign (MJD 54850-54983) and the panel {\em (c)} for
    the period after the campaign (MJD54983-55248). In all panels, 
     the black line depicts the result of the unbinned likelihood PL
     fit and  the red contours denote the $68\%$ uncertainty of the PL
     fit.  The legend reports the results from the unbinned likelihood PL fit in the energy range $0.3-400$\,GeV.}
  \label{fig:SED_LongIntervals}
\end{figure}

The Mrk\,421 spectrum measured by \FermiLAT during the period covered
by the multifrequency campaign is shown in panel {\em (b)} of
Fig\,\ref{fig:SED_LongIntervals}. The spectrum can be described with a
single PL function with photon index $1.75 \pm 0.03$ and photon flux
$F(>0.3$ GeV$) = (6.1 \pm 0.3) \times 10^{-8}$\,ph\,cm$^{-2}$s$^{-1}$;
which is somewhat lower than the average spectrum over the first 1.5 years of \FermiLAT operation (see Figure\,\ref{fig:SED}).

For comparison purposes we also computed the spectra for the time
periods before and after the multifrequency campaign (the time
intervals MJD 54683-54850 and MJD 54983-55248, respectively). These two
spectra are shown in panels {\em (a)} and {\em (c)} of
Figure\,\ref{fig:SED_LongIntervals}. The two spectra can be described
very satisfactorily with single PL functions of photon indices $1.79 \pm
0.03$ and $1.78 \pm 0.02$ and photon fluxes $F(>0.3$~GeV$) = (7.1 \pm
0.3) \times 10^{-8}$\,ph\,cm$^{-2}$s$^{-1}$ and $F(>0.3$~GeV$) = (7.9 \pm
0.2) \times 10^{-8}$\,ph\,cm$^{-2}$s$^{-1}$. Therefore, during the
multifrequency campaign, Mrk\,421 showed a spectral shape that is
compatible with the periods before and after the campaign, and a
photon flux which is about 20\% lower than before the campaign and
30\% lower than after the campaign.

\subsection{The Average Broad Band SED during the Multifrequency Campaign}
\label{MWSEDDataResults}

\begin{figure}[t]
 \centering
  \includegraphics[width=6.0 in]{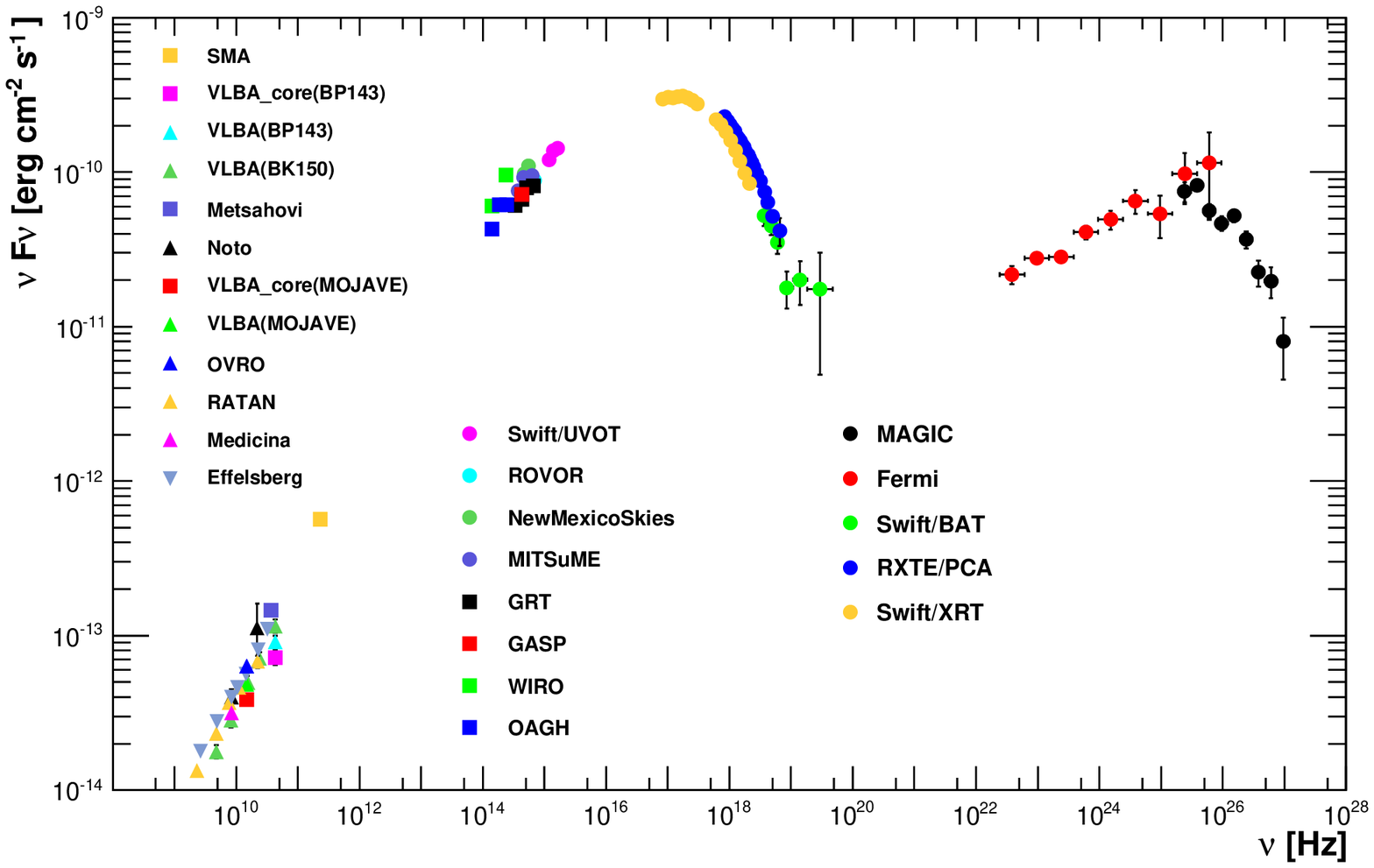}
  \caption{Spectral energy distribution of Mrk\,421 averaged over all
    the observations taken during the multifrequency campaign from
    2009 January 19 (MJD 54850) to 2009 June 1 (MJD 54983). The legend reports the correspondence between the instruments and the measured fluxes.  The host galaxy has been subtracted, and the optical/X-ray data were corrected 
for the Galactic extinction. The TeV data from MAGIC were corrected
for the absorption in the extragalactic background light using the
prescription given in \cite{franc08}.}
  \label{fig:MWSED}
 \end{figure}

The average SED of Mrk\,421 resulting from our 4.5-month-long
multifrequency campaign is shown in Figure\,\ref{fig:MWSED}.  This is
the most complete SED ever collected for Mrk\,421 or for any other BL
Lac object (although an SED of nearly similar quality was reported in
\cite{AbdoMrk501} for Mrk 501).  At the highest energies, the
combination of \FermiLAT and MAGIC allows us to measure, for the first
time, the high energy bump without any gap; both the rising and
falling segments of the components are precisely determined by the
data.  The low energy bump is also very well measured; \Swiftc/BAT and
\RXTEc/PCA describe its falling part, \Swiftc/XRT the peak, and the
\Swiftc/UV and the various optical and IR observations describe the
rising part.  The rising tail of this peak was also measured with
various radio instruments. Especially important are the observations
from SMA at 225\,GHz, which help connecting the the bottom (radio) 
to the peak (optical/X-rays) of the synchrotron bump (in the $\nu F_{\nu}$ representation).
The flux
measurements by VLBA, especially the ones corresponding to the core,
provide us with the radio flux density from a region that is
presumably not much larger than the blazar emission region. Therefore,
the radio
flux densities from interferometric observations (from the VLBA core)
are expected to be close upper limits to the radio continuum of the
blazar emission component. On the other hand, the low frequency radio
observations performed with single dish instruments have a relatively
large contamination from the non-blazar emission and are probably
considerably above the energy flux from the blazar emission region.
{\bf  The only spectral intervals lacking observations are from 1 meV -
0.4eV, and  200 keV - 100 MeV,
 where the sensitivity of the current
instruments is insufficient to detect Mrk\,421.  We note however,
that the detailed GeV coverage together with our
broadband, 1-zone SSC modeling strongly constrains the expected
emission in the difficult to access 1 meV - 0.4 eV bandpass.}

{\bf During this campaign, Mrk\,421 showed low activity and relatively
small flux variations at all frequencies \citep{PanequeComo2009}.
At VHE ($>$ 100 GeV), the measured flux is half the flux from the Crab Nebula, which
is among the lowest fluxes recorded by MAGIC for this source 
\citep{Mrk421MAGIC,AleksicMrk4212010}. At X-rays, the fluxes observed
during this campaign are about 15 mCrab, which is about 3 times
higher than the lowest fluxes measured by \RXTEc/ASM since 1996.
Therefore, because of the low flux, low (multifrequency) variability and
the large density of observations, the collected data during this
campaign can be considered an excellent proxy for the low/quiescent
state SED of Mrk\,421. } It is worth stressing that the good agreement
in the overlapping energies of the various instruments (which had
somewhat different time coverage during the campaign) supports this
hypothesis.

%% ============================================================================
%%
%% SECTION 6 -- Model fitting 
%%
%% ============================================================================

\section{SED Modeling} 
\label{SEDModel}

{\bf We turn now to modeling the multifrequency data set collected during
the 4.5 month campaign in the context of homogeneous hadronic and
leptonic models. The models discussed below assume emission
mainly from a single, spherical and homogeneous region of the jet.
This is a good approximation to model flaring events 
with observed correlated variability (where the dynamical time scale does not exceed the 
flaring time scale significantly),  although it is an
over-simplification for quiescent states, where 
the measured blazar emission might be produced by the radiation from 
different zones characterized by different values of the relevant
parameters.  There are
several models in the literature along those lines 
\citep[e.g.][]{ghisellini05,Katar2008,graff08,Giannios2009}
 but at the cost of introducing more free
parameters that are, consequently, less well constrained and more
difficult to compare between models.  This is particularly problematic
if a ``limited'' data set (in a 
time and energy coverage) is employed in the modeling,
although it could work well if the amount of multifrequency data is
extensive enough to substantially constrain the parameter space.
In this work we adopted the 1-zone homogeneous models for their
simplicity as well as for being able to compare with previous works. 
The 1-zone homogeneous models are the most widely used models to describe the SED of high-peaked BL Lacs.
Furthermore, although the modeled SED is averaged over 4.5 months of
observations, the very low observed multifrequency variability during this campaign, and
in particular the lack of strong keV and GeV variability (see Figs \ref{fig:Lc7days} and \ref{fig:Lc7daysMW}) in these timescales
suggests that the presented data are a good representation of the average
broad-band emission of Mrk\,421 on timescales of few days. We, therefore,
feel confident that the physical parameters required by our modeling to
reproduce the average 4.5 month SED are a good representation of the
physical conditions at the emission region down to timescales of a few
days, which is comparable to the dynamical timescale derived from the
models we discuss. The implications (and caveats) of the modeling
results are discussed in section \ref{Discussion}.}

Mrk\,421 is at a relatively low redshift
($z=0.031$), yet the attenuation of its VHE MAGIC spectrum by the
extragalactic background light (EBL) is non-negligible for all models
and hence needs to be accounted for using a parameterization for the
EBL density.  The EBL absorption at 4~TeV, the highest energy bin of
the MAGIC data (absorption will be less at lower energies), varies
according to the model used from $e^{-\tau_{\gamma\gamma}}=0.29$ for
the ``Fast Evolution'' model \cite[]{stecker06} to
$e^{-\tau_{\gamma\gamma}}=0.58$ for the models of \cite{franc08} and
\citet{gilmore09}, with most models giving
$e^{-\tau_{\gamma\gamma}}\sim 0.5$--0.6, including the model of
\cite{finke10} and the ``best fit'' model of \cite{kneiske04}.  We
have de-absorbed the TeV data from MAGIC with the \cite{franc08}
model, although most other models give comparable results.

\subsection{Hadronic Model}
\label{hadronicmodelSEDText}

\begin{figure}
\epsscale{1.0}
\plotone{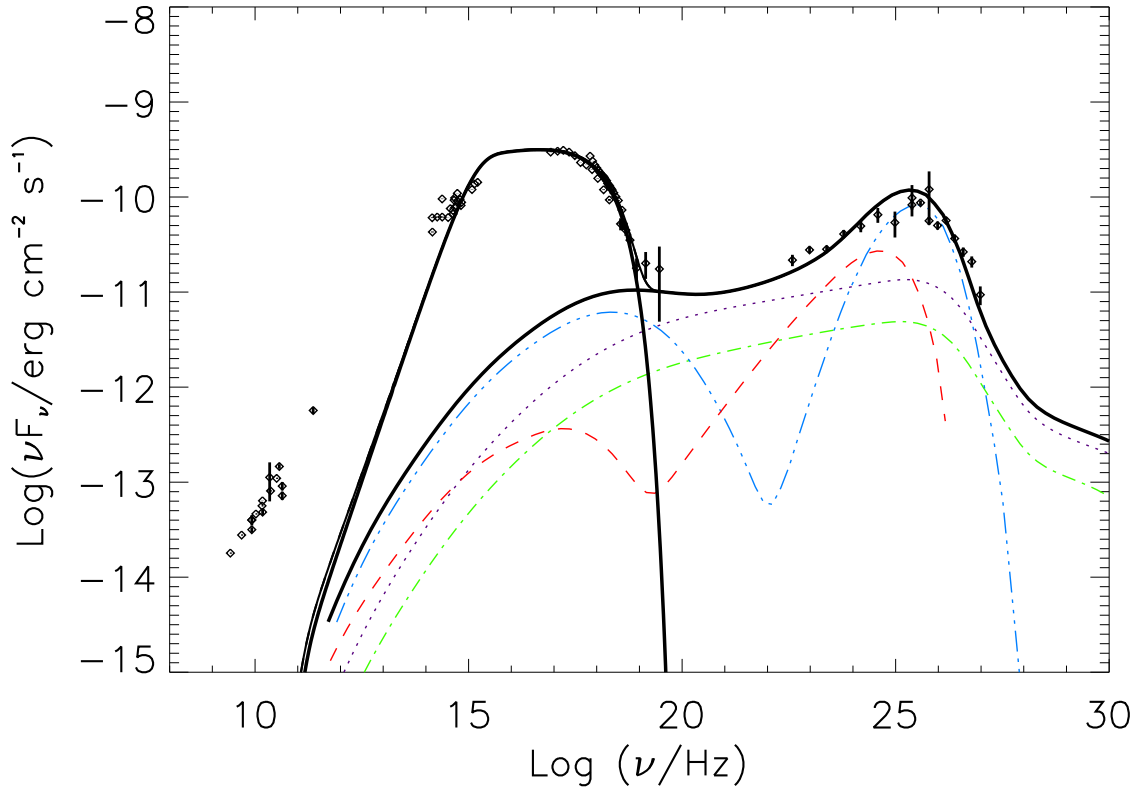}
\caption{ Hadronic model fit components: $\pi^0$-cascade (black dotted
line), $\pi^\pm$ cascade (green dashed-dotted line), $\mu$-synchrotron and
cascade (blue dashed-triple-dotted line), proton synchrotron and cascade
(red dashed line). The black thick solid line is the sum of all
emission components (which also includes the
synchrotron emission of the primary electrons at optical/X-ray frequencies). 
The resulting model parameters are reported in table
\ref{modelparamsHadron}.
}
\label{hadronicmodelSED}
\end{figure}

\begin{figure}
\epsscale{1.0}
\plotone{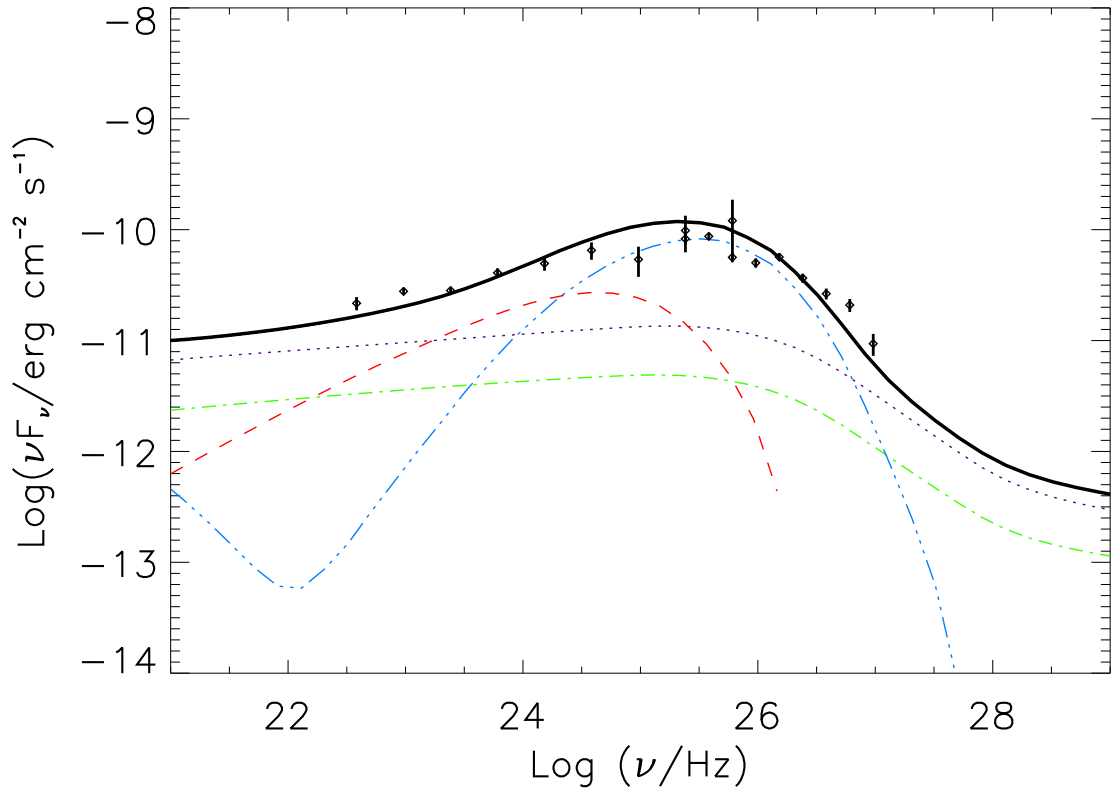}
\caption{Expanded view of the high energy bump of the SED data and model
  presented in Figure~\ref{hadronicmodelSED}.
}
\label{hadronicmodelSEDZoom}
\end{figure}

\begin{deluxetable}{lcc}
\tabletypesize{\scriptsize}
\tablecaption{
Parameter values from the SPB model fit to the SED from
Mrk\,421 shown in Figure~\ref{hadronicmodelSED}.
}
\tablewidth{0pt}
\tablehead{
\colhead{Parameter} &
\colhead{Symbol} &
\colhead{Value} 
}
\startdata
Doppler Factor & $\delta$      &  12\\
Magnetic Field [G] & $B$ & $50$ \\
Comoving blob radius [cm] & $R$ & $4\times10^{14}$ \\
\hline
Power law Index of the injected electron distribution $^{a}$& $\alpha_e$ & 1.9 \\
Power law Index of the injected proton distribution $^{a}$ & $\alpha_p$ & 1.9 \\
Minimum Electron Lorentz Factor & $\gamma_{e,min}$ & $7 \times10^2$ \\
Maximum Electron Lorentz Factor & $\gamma_{e,max}$ & $4 \times10^4$ \\
Minimum Proton Lorentz Factor $^{b}$& $\gamma_{p,min}$ & $1$ \\
Maximum Proton Lorentz Factor & $\gamma_{p,max}$ & $2.3 \times10^9$ \\
Energy density in protons [erg cm$^{-3}$] & $u\prime_p$ & $510$ \\
Ratio of number of electrons with respect to protons & $e/p$ & $90$ \\
\hline
Jet Power [erg s$^{-1}$] &  $P_{jet} $  &   $4.5\times 10^{44}$\\
\enddata
\tablecomments{$(a)$ The model assumes $\alpha_e=\alpha_p$, hence only
one free parameter.}
\tablecomments{$(b)$ The parameter $\gamma_{p,min}$ was fixed to the
    lowest possible value, 1, and hence this is actually not a free parameter.}
\label{modelparamsHadron}
\end{deluxetable}

 If relativistic protons are present in the jet of Mrk~421, hadronic
interactions, if above the interaction threshold, must be considered
for modeling the source emission. For the present modeling we use the
hadronic Synchrotron-Proton Blazar (SPB) model of
\cite{muecke01,muecke03}.  Here, the relativistic electrons (e)
injected in the strongly magnetized (with {\bf homogeneous} magnetic field
with strength $B$) blob lose energy predominantly through
synchrotron emission. The resulting synchrotron radiation of the
primary e component dominates the low energy bump of the blazar SED,
and serves as target photon field for interactions with the
instantaneously injected relativistic protons (with index
$\alpha_p=\alpha_e$) and pair (synchrotron-supported) cascading.

Figures \ref{hadronicmodelSED} and \ref{hadronicmodelSEDZoom} show a
satisfactory (single zone) SPB model representation of the data from 
Mrk\,421 collected during the campaign. The corresponding parameter
values are reported in Table \ref{modelparamsHadron}.  In order to fit
the optical data, the lowest energy of the injected electrons is 
required to be maintained as $\gamma_{e,min}\approx 700$ 
through the steady state. This requires {\bf a continuous electron
injection rate density of at least $\gapp 1.4$cm$^{-3}$s$^{-1}$} to
balance the synchrotron losses at that energy, and is about a factor
$\sim 100$ larger than the proton injection rate.  The radio fluxes
predicted by the model are significantly below the observed 8-230\,GHz radio
fluxes. This is related to the model being designed to follow the
evolution of the jet emission during $\gamma$-ray production where
radiative cooling dominates over adiabatic cooling.  Here, the
emission region is optically thick up to $\sim$100\,GHz frequencies,
and the synchrotron cooling break ($\gamma_e\sim 10$) would be below
the synchrotron-self-absorption turn-over.  The introduction of
additional, poorly constrained components would be necessary to
account for the subsequent evolution of the jet through the expansion
phase where the synchrotron radiation becomes gradually optically thin
at cm wavelengths. This is omitted in the following modeling.

The measured spectra in the $\gamma$-ray band ($>1$ GeV) is dominated
by synchrotron radiation from short-lived muons (produced during
photomeson production) as well as proton synchrotron radiation, with
significant overall reprocessing, while below this energy the
$\pi$-cascade dominates.  The interplay between muon and proton
synchrotron radiation together with appreciable cascade synchrotron
radiation initiated by the pairs and high energy photons from
photomeson production, is responsible for the observed MeV-GeV flux.
The TeV emission is dominated by the high energy photons from the muon
synchrotron component. The source intrinsic model SED predicts
$>10$ TeV emission on a level of 2 to 3 orders of magnitude below the
sub-TeV flux, which, will be further weakened by $\gamma$-ray
absorption by the EBL.

{\bf The overall required particle and field energy 
density are within a factor 5 of equipartition, and
a total jet power (as measured in the galaxy rest frame) of $4\times 
10^{44}$ erg s$^{-1}$ in agreement with expectations
for a weakly accreting disk of a BL Lac object \citep[see][]{Cao2003}.}

Alternative model fits are possible if the injected electron and
proton components do not have the same power-law index. This
"relaxation" of the model would add one extra parameter and so would
allow for improvement in the data-model agreement, especially around
the synchrotron peak and the high energy bump.  It would also allow a
larger tolerance on the size region R, which is considered to be small
in the SPB model fit presented here.

\subsection{Leptonic Model}
\label{leptonicmodel}

The simplest leptonic model typically used to describe the emission
from BL Lac objects is the 1-zone Synchrotron Self-Compton model
(SSC).  Within this framework, the radio through X-ray emission is
produced by synchrotron radiation from electrons in a homogeneous,
randomly-oriented magnetic field ($B$) and the the $\gamma$-rays are
produced by inverse Compton scattering of the synchrotron photons by
the same electrons which produce them. For this purpose, we use the
1-zone SSC code described in \citet{finke08}.  The electron
distribution from 1-zone SSC models is typically parameterized with
one or two power-law (PL) functions (that is, zero or one break)
within an electron Lorentz factor range defined by $\gamma_{min}$ and
$\gamma_{max}$ (where the electron energy is $\gamma m_e c^2$).  We
use the same approach in this work. However, we find that, in order to
properly describe the shape of the measured broad-band SED during the
4.5 months long campaign, the model requires an electron distribution
parameterized with three PL functions (and hence two breaks). In other
words, we must add 2 extra free parameters to the model: the second
break at $\gamma_{brk,2}$ and the index of the third PL function
$p_3$. Note that a second break was also needed to describe the SED of
Mrk 501 in the context of the synchrotron/SSC model
\citep{AbdoMrk501}.  An alternative possibility might be to use an
electron distribution parametrized with a curved function such as that
resulting from episodic particle acceleration \citep{Perlman05} or the
log-parabolic function used in \citet{Tramacere2009}. However, we note
that such a parameterization might have problems describing the
highest X-ray energies, where the current SED data (\RXTEc/PCA and
\Swiftc/BAT) do not show a large spectrum curvature.

Even though the very complete SED constrains the shape of the electron
distribution quite well, there is still some degeneracy in the range
of allowed values for the general source parameters $R$ (comoving blob
radius), $B$ and $\delta$ (doppler factor).  For a given break in the
measured low energy (synchrotron) bump, the break in the electron
distribution $\gamma_{brk}$ scales as $1/\sqrt{B \delta}$. In order to
minimize the range of possible parameters, we note that the emitting
region radius is constrained by the variability time, $t_v$, so that
\begin{equation}
\label{Rb_eqn}
R = \frac{\delta c t_{v,min}}{1+z} \le \frac{\delta c t_{v}}{1+z}\ .
\end{equation}

%We note here that the above mentioned expression is valid only under the assumption of a 
%region with spherical geometry where the radiative loss timescale $t_{\rm loss}$ is larger than the crossing timescale $t_{\rm cross}$.
%In the cases where $t_{\rm loss} < t_{\rm cross}$,  the particles are in strong cooling regime and 
%hence they radiate their energy (and hence produce flux variations) on distances smaller than $R_b$
%(and hence show flux variability timescales smaller than $t_{v,min}$). 

\begin{figure}[t]
\epsscale{1.0}
\plotone{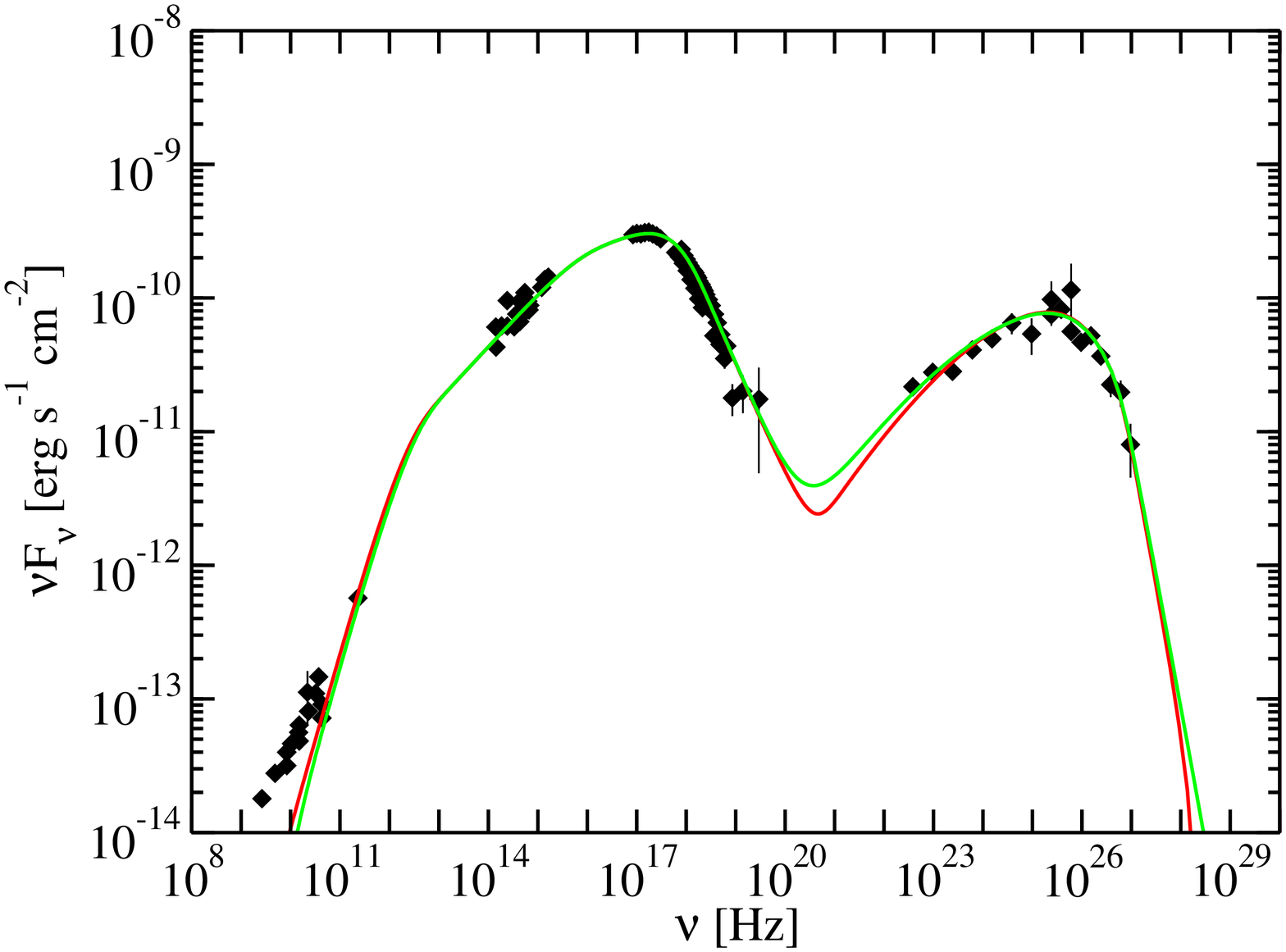}
\caption{ SED of Mrk\,421 with two 1-zone SSC model fits obtained with
different minimum variability timescales: $t_{var} = 1~$day (red
curve) and $t_{var} = 1~$hour (green curve) .  The parameter values
are reported in Table \ref{modelparams}.  See text for further
details.}
\label{modelSED}
\end{figure}

During the observing campaign, Mrk\,421 was in a rather low activity
state, with multifrequency flux variations occurring on timescales
larger than 1 day \citep{PanequeComo2009}, so we used $t_{v,min} = 1$
day in our modeling. In addition, given that this only gives an upper
limit on the size scale, and the history of fast variability detected
for this object \citep[e.g.][]{Gaidos1996,giebels07}, we also
performed the SED model using $t_{v,min} = 1$ hour. The resulting SED
models obtained with these two variability timescales are shown in
Figure~\ref{modelSED}, with the parameter values reported in table
\ref{modelparams}.  The blob radii are large enough in these models
that synchrotron self-absorption (SSA) is not important; for the
$t_{v,min}=1$ hr model, $\nu_{SSA}=3\times10^{10}$ Hz, at which
frequency a break is barely visible in Figure~\ref{modelSED}.  It is
worth stressing the good agreement between the model and the data: the
model describes very satisfactorily the entire measured broad-band
SED. The model goes through the SMA (225\,GHz) data point, as well as
through the VLBA (43\,GHz) data point for the partially-resolved radio
core. The size of the VLBA core of the 2009 data from Mrk\,421 at
15\,GHz and 43\,GHz is $\simeq$ 0.06--0.12\,mas\, (as reported in 
Section \ref{radio}) or, using the
conversion scale 0.61\,pc/mas $\simeq$ 1--2 $\times 10^{17}$\,cm.  The
VLBA size estimation is the FWHM of a Gaussian representing the
brightness distribution of the blob, which could be approximated as
0.9 times the radius of a corresponding spherical blob
\citep{Marscher1983}.  That implies that the size of the VLBA core is
comparable (factor $\sim$2--4 larger) than that of the model blob for
$t_{var} = 1~$day ($\sim 5 \times 10^{16}$ cm). Therefore, it is
reasonable to consider that the radio flux density from the VLBA core
is indeed dominated by the radio flux density of the blazar
emission. 
%Forthcoming multi-band correlation studies (in particular
%VLBA and SMA radio with the $\gamma$-rays from \FermiLATc) will shed
%light on this particular subject. 
The other radio observations are
single dish measurements and hence integrate over a region that is
orders of magnitude larger than the blazar emission. Consequently, we
treat them as upper limits for the model.

{\bf The powers of the different jet components derived from the model fits (assuming $\Gamma = \delta$) are also reported in Table
\ref{modelparams}.  Estimates for the mass of the supermassive black
hole in Mrk\,421 range from $2\times10^8\ M_\odot$ to $9\times10^8\
M_\odot$ \citep{barth03,wu02}, and hence the Eddington luminosity should
be between $2.6 \times 10^{46}$ and $1.2 \times 10^{47}$ erg s$^{-1}$, 
that is well above the jet luminosity.}

\begin{deluxetable}{lccc}
%\rotate
\tabletypesize{\scriptsize}
\tablecaption{
Parameter values from the 1-zone SSC model fits to the SED from
Mrk\,421 shown in Figure~\ref{modelSED}.
}

\tablewidth{0pt}
\tablehead{
\colhead{Parameter} &
\colhead{Symbol} &
\colhead{Red Curve} &
\colhead{Green Curve}
}
\startdata
Variability Timescale [s]$^{a}$ & $t_{v,min}$ & $8.64\times10^4$ & $3.6\times10^3$\\
Doppler Factor & $\delta$      &  21 & 50 \\
Magnetic Field [G] & $B$ & $3.8 \times 10^{-2}$ & $8.2\times10^{-2}$\\
Comoving blob radius [cm] & $R$ & $5.2\times10^{16}$ & $5.3\times10^{15}$ \\
\hline
Low-Energy Electron Spectral Index & $p_1$       & 2.2 & 2.2\\
Medium-Energy Electron Spectral Index  & $p_2$       & 2.7 & 2.7 \\
High-Energy Electron Spectral Index  & $p_3$       & 4.7 & 4.7 \\
Minimum Electron Lorentz Factor & $\gamma_{min}$ & $8.0\times10^2$ & $4\times10^2$ \\
Break1 Electron Lorentz Factor & $\gamma_{brk1}$ & $5.0\times10^4$ & $2.2\times10^4$ \\ 
Break2 Electron Lorentz Factor & $\gamma_{brk2}$ & $3.9\times10^5$ & $1.7\times10^5$\\ 
Maximum Electron Lorentz Factor & $\gamma_{max}$  & $1.0\times10^8$ & $1.0\times10^8$ \\
\hline
Jet Power in Magnetic Field [erg s$^{-1}$]$^{b}$ & $P_{j,B}$  & $1.3\times10^{43}$ & $3.6\times10^{42}$\\
Jet Power in Electrons [erg s$^{-1}$] & $P_{j,e}$ & $1.3\times10^{44}$ & $1.0\times10^{44}$\\
Jet Power in Photons [erg s$^{-1}$]$^{b}$ & $P_{j,ph}$ & $6.3\times10^{42}$ & $1.1\times10^{42}$\\ 
\enddata
\tablecomments{$(a)$ The variability timescale was not derived from
  the model fit, but rather used as an input (constrain) to the
  model. See text for further details. }
\tablecomments{$(b)$ The quantities $P_{j,B}$  and $P_{j,ph}$ are
  derived quantities; only $P_{j,e}$ is a free parameter in the model.} 
\label{modelparams}
\end{deluxetable}

It is important to note that the parameters resulting from the
modeling of our broad-band SED differ somewhat from the parameters
obtained for this source of previous works
\citep{Krawczynski2001,Blazejowski2005,Mrk421Whipple2006, albert07,
giebels07, fossati08, finke08, Horan2009, Acciari2009}.  One
difference, as already noted, is that an extra break is required.
This could be a feature of Mrk~421 in all states, but we only now have
the simultaneous high quality spectral coverage to identify it.  For
the model with $t_{var}= 1~$day (which is the time variability
observed during the multifrequency campaign), additional differences
with previous models are in $R$, which is an order of magnitude
larger, and $B$, which is an order of magnitude smaller.  This 
mostly results from the longer variability time in this low state.
Note that using a shorter variability ($t_{var}= 1~$hour, green curve)
gives a smaller $R$ and bigger $B$ than most models of this source.

\begin{figure}[t]
\epsscale{1.0}
\plotone{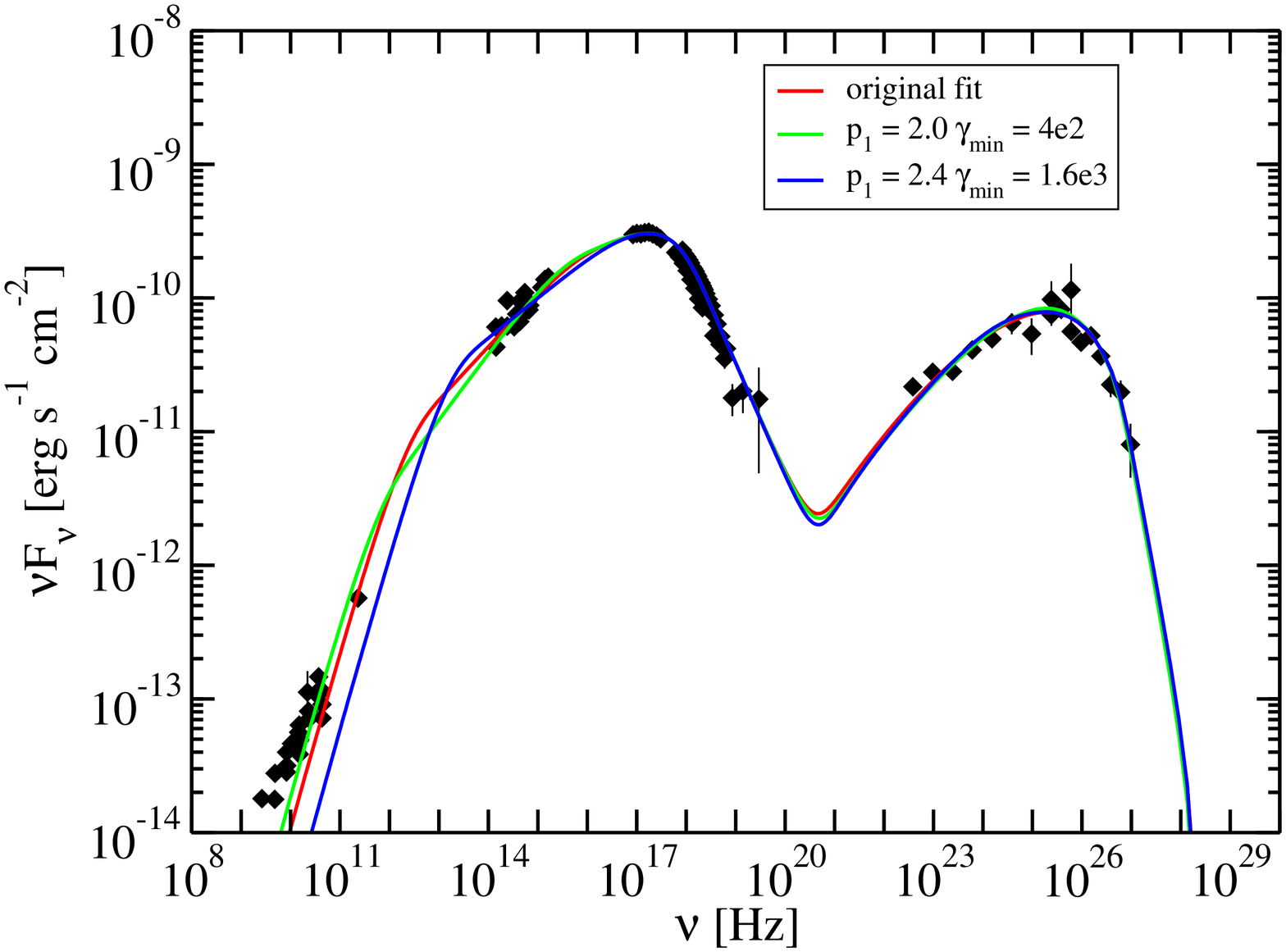}
\caption{ SSC model fit of the SED from Mrk\,421 presented in
  Figure~\ref{modelSED} (for $t_{var} \sim $1 day), with
  variations by a factor of 2 of the parameter $\gamma_{min}$,
  together with adjustments in the parameter $p_1$ in order to match
  the experimental data.  See text for further
details.}
\label{modelSEDGammaMin}
\end{figure}

Another difference in our 1-zone SSC model with respect to previous
works relate to the parameter $\gamma_{min}$. This parameter has
typically not been well constrained because the single-dish radio data can only be
used as upper limits for the radio flux from the blazar emission. This
means that the obtained value for $\gamma_{min}$ (for a given set of
other parameters $R$, $B$, and $\delta$) can only be taken as a lower
limit: a higher value of $\gamma_{min}$ is usually possible. {\bf  In our
modeling we use simultaneous \FermiLAT data as well as  SMA
and VLBA radio data, which we assume is dominated by the blazar
emission. We note that the size of the emission from our SED model fit 
(when using $t_{var} \sim$1 day) is comparable to the partially resolved VLBA radio core and hence we think this assumption is reasonable.
The requirement that the model SED fit goes through those
radio points puts further constrains to the model, and in particular to the parameter $\gamma_{min}$: a decrease in the value of $\gamma_{min}$ would over-predict the radio
data, while an increase of $\gamma_{min}$ would under-predict the SMA
and VLBA core radio data, as well as the \FermiLAT spectrum below 1
GeV if the increase in $\gamma_{min}$ would be large. We explored model fits with different $\gamma_{min}$ and $p_1$, and
found that, for the SSC
model fit with $t_{var} = 1~day$ (red curve in Figure~\ref{modelSED}), $\gamma_{min}$ is well constrained within a factor of 2 to the
value of $8 \times 10^2$ (see Figure~\ref{modelSEDGammaMin}). In the case of the SSC model with $t_{var}$
= 1 hour (green curve in Figure~\ref{modelSED}), if we make the same
assumption that the SMA and VLBA core emission is dominated by the
blazer emission\footnote{In the case of $t_{var} \sim$ 1 hour, the
  size of the emission region derived from the SSC model is one order
  of magnitude smaller than the size of the VLBA core and hence the
  used assumption is somewhat less valid than for the model with
  $t_{var} \sim 1$ day}, $\gamma_{min}$ can be between 
$2 \times 10^2$ up to $10^3$, and still provide a good match to the
SMA/VLBA/optical data and the \FermiLAT spectrum.
In any case, for any variability timescale, the electron distribution
does not extend down to $\gamma_{min} \sim 1$ to a few, and is constrained within a factor of 2.
This is particularly
relevant because, for power-law distributions with index p $>$ 2, the
jet power carried by the electrons is dominated by the low energy
electrons.} Therefore, the tight constraints on $\gamma_{min}$
translate into tight constraints on the jet power carried by the
electrons. For instance, in the case of the model with $t_{var} =
1~$hour, using $\gamma_{min}=10^{3}$ (instead of $\gamma_{min}=4
\times 10^{2}$) would reduce the jet power carried by electrons from
$P_{j,e}\approx 10^{44}$ erg s$^{-1}$ down to $P_{j,e}\approx
8\times10^{43}$ erg s$^{-1}$.

Another parameter where the results presented here differ from
previous results in the literature is the first PL index $p_1$.  This
parameter is dominated by the optical and UV data points connecting
with the \Swiftc/XRT, as well as by the necessity of matching the
model with the \FermiLAT GeV data.  Note that our model fit also goes
over the SMA and VLBA (partially-resolved) core fluxes. Again, since
these constrains did not exist (or were not used) in the past, most of
the 1-zone SSC model results (for Mrk\,421) in the literature report a
$p_1$ value that differs from the one reported in this work. We note
however that the values for the parameters $p_2$ and $p_3$ from our
model fits, which are constrained mostly by the X-ray/TeV data, are
actually quite similar to the parameters $p_1$ and $p_2$ from the
previous 1-zone SSC model fits to Mrk\,421 data.

%% ============================================================================
%%
%% SECTION 7 -- Discussion 
%%
%% ============================================================================

\section{Discussion}
\label{Discussion}

{\bf In this section of the paper we discuss the implications of the
experimental and SED modeling results presented in the previous
sections. As explained at the beginning of section 6, 
for simplicity and for the sake of comparison with previously 
published results, we modeled the SED with scenarios
based on 1-zone homogeneous emitting regions, which are commonly 
used to parameterize the broad-band emission in blazars. 
We note that this is a simplification of the problem; the 
emission in blazar jets could be produced in an inhomogeneous or
stratified region, as well as in N independent regions. 
An alternative and quite realistic scenario could be a  standing shock where
particle acceleration takes place and radiation is being produced
as the jet flow or superluminal knots cross it
\citep[e.g.][]{Komissarov1997, Marscher2008}. The Lorentz factor of the plasma, as it flows through the standing (and by necessity
oblique) shock  is the Lorentz factor (and through setting the angle, the 
Doppler factor) of the model. We note however that, as discussed in \citet{Sikora1997}, 
the steady-sate emission could also be parametrized by N moving blobs
that only radiate when passing through the standing shock. If at any
given moment, only one of these blobs were visible at the observer frame, the 1-zone
homogeneous model could be a plausible approximation of the standing-shock scenario. 

In any case, the important thing is that, 
in the proposed physical scenario, the stability time-scale of the particle accelerating shock front
is not connected to the much shorter cooling times that give rise to
spectral features.  For as long as the injection of particles in the blob and the dynamics of
the blob remain unchanged, the SED, along with the breaks due to
radiative cooling and due to the value of $\gamma_{min}$ where Fermi
acceleration  presumably picks up, will remain unchanged. The lack of
(substantial) multifrequency variability observed during this campaign
suggests that this is the case, and hence that the 4.5-months-averaged
SED is also representative of the broad-band emission of SED during
much shorter periods of time that are comparable to the dynamical timescales derived from
the models.}

\subsection{What are the Spectral  Breaks Telling Us?}
\label{breaks}

%We examine now  the possibility that the observed SED is not produced in a single location with
%homogenous physical characteristics. Although a detailed study of this possibility requires an in-depth analysis and inhomogenous modeling (e.g. Graff et al. 2008) of the multiwavelength variability of the source,  certain conclusions can be drown from  modeling  the observed time-averaged SED.  
%An important feature of the SED is  that
% above the $\nu f_\nu$
%peaks of both the low and high energy components we see the emission  not exhibiting a cutoff, but continuing with steep power laws of similar slope for both the X-ray and TeV bands (spectral index $\alpha\approx 1.8$). In the context of the leptonic model, this means the electron energy distribution ( ) continues with a power law form at energies higher than those responsible for producing the  $\nu f_\nu$ peaks of both components. 

In our homogenous leptonic model we reproduce the location of the $\nu
f_\nu$ peaks by fitting the Lorentz factors $\gamma_{brk,1}$ and
$\gamma_{brk,2}$ (as well as the values of $B$ and $\delta$) where the
electron energy distribution breaks. There is, however, a Lorentz
factor where one typically (in blazar modeling) expects a break in the
electron energy distribution (EED), and this is the Lorentz factor
$\gamma_c=3 \pi m_e c^2/(\sigma_\tau B^2 R)$ where the escape time
from the source equals the radiative (synchrotron) cooling time.  The
fact that the values of the second break, $\gamma_{brk,2}$, fit by our
leptonic models $(\gamma_{brk,2}=3.9 \times 10^5, 1.7 \times 10^5)$
are similar to the Lorentz factors $(\gamma_c=1.6 \times 10^5, 3.3
\times 10^5$) where a cooling break in the EED is expected, strongly
suggests that the second break in the EED derived from the modeling is
indeed the cooling break.

%{\bf Markos, I did change the numbers according to the latest fits
%from Justin. Check that those numbers are correct. It is interesting
%to see that, even though the values match within a factor 2, the
%measured/modeled breaks and cooling-predicted breaks evolve in
%opposite directions when going from tvar=1day to tvar=1hour. TO BE
%CHECKED.}

The observed spectral shape in both the low and high energy SED
components are reproduced in our homogenous model  by a change of
electron index $\Delta p = p_3 - p_2=2.0$.  Such a large
break in the EED is in contrast to the canonical cooling break $\Delta
p = p_3 - p_2=1.0$ that produces a spectral index change of $\Delta
\alpha=0.5$, as predicted for homogenous models (e.g., Longair
1994). An attempt to model the data fixing $\Delta p = p_3 - p_2=1.0$
gave unsatisfactory results, and hence this is not an option; a large
spectral break is needed.  It would be tempting to speculate that what
we observe is not a cooling break, but rather something that results
from a characteristic of the acceleration process which is not
understood and that, therefore, does not bind us to the $\Delta p =
1.0$ constraint. But we would then have to attribute to shear fortuity
the fact that the Lorentz factors where this break takes place are
very close to the Lorentz factors where cooling is actually expected.
 
The question that naturally arises is why, although the EED break
postulated by the homogeneous model is at nearly the same energy as
the expected cooling break, the spectral break observed is stronger.
Such strong breaks are the rule rather than the exception in some non-thermal
sources like pulsar-wind nebulae and extragalactic jets
\citep[see][]{Reynolds09} and the explanations that have been given
relax the assumption of a homogenous emitting zone, invoking gradients
in the physical quantities describing the system
\citep{Marscher1980}. In all inhomogenous models, electrons are
injected at an inlet and are advected downstream suffering radiative
losses that result in the effective size of the source declining with
increasing frequency for a given spectral component.  In sources where
the beaming of the emitted radiation is the same throughout the source
(this is the case for non-relativistic flows or for relativistic flows
with small velocity gradients), the spectral break formed is stronger
than the canonical $\Delta \alpha=0.5$ if the physical conditions
change in such a way that the emissivity at a given frequency
increases downstream (Wilson 1975, Coleman \& Bicknell 1988, Reynolds
2009).

If, in addition to these considerations, we allow for significant
relativistic velocity gradients, either in the form of a decelerating
flow \citep{georganopoulos03} or the form of a fast spine and slow sheath
flow \citep{ghisellini05}, the resulting differential beaming of the
emitted radiation can result in spectral breaks stronger that $\Delta
\alpha\approx 0.5$. Studies of the SEDs of sources with different jet
orientations (e.g. radio galaxies and blazars) can help to understand
the importance of differential beaming, and therefore of relativistic
velocity gradients in these flows.  Because in all these models the
volume of the source emitting at a given frequency is connected to the
predicted spectral break, it should be possible to use the variability
time-scale at different frequencies to constrain the physics of the
inhomogenous flow.
 
\subsection{Physical Properties of Mrk~421}
\label{Mrk421Properties}

As mentioned in \S \ref{MWSEDDataResults}, the SED emerging from the
multifrequency campaign is the most complete and accurate 
representation of the low/quiescent state of Mrk\,421 to date. This
data provided us with an unprecedented opportunity to constrain and tune
state-of-the-art modeling codes.  In \S\ref{SEDModel} we modeled the SED within
two different frameworks: a leptonic and a hadronic scenario. Both
models are able to represent the overall SED. As can be seen in Figures 
\ref{hadronicmodelSED} and \ref{modelSED}, the leptonic model fits describe
the observational data somewhat better than the hadronic model; yet 
we also note here that, in this paper, the leptonic model has one more
free parameter than the hadronic model. A very efficient way of 
discriminating between the two scenarios would be through 
multi-wavelength variability observations. 
%This topic will be
%discussed in the forthcoming publication with the multifrequency
%variability/correlations from the data we collected during the
%observing campaign.  
It is however interesting to discuss the
differences between the two model descriptions we presented above.

\subsubsection{Size and location of the emitting region}

The characteristic size to which the size of the emitting region must
be compared is the gravitational radius of the Mrk 421 black hole. For
a black hole mass of $\sim 2-9 \times 10^{8} M_\odot$ \citep{barth03,
wu02}, the corresponding size is $R_g\approx 0.5-2.0 \times 10^{14}$
cm.  In the hadronic model the source size can be as small as $R =
4\times 10^{14}$ cm (larger source sizes can not be ruled out though;
see Sect.~6.1), within one order of magnitude of the gravitational
radius. The consequence is a dense synchrotron photon energy density
that facilitates frequent interactions with relativistic protons,
resulting in a strong reprocessed/cascade component which leads to a
softening of the spectrum occurring mostly below 100 MeV.  The
Fermi-LAT analysis presented in this paper (which used the instrument 
response function given by \texttt{P6\_V3\_DIFFUSE}) is not sensitive to
these low energies and hence the evaluation of this potential
softening in the spectrum will have to be done with future analyses
(and more data).  This will potentially allow accurate determination
of spectra down to photon energies of $\sim 20$ MeV with the LAT.

The leptonic model can accommodate a large range of values for $R$, as
long as it is not so compact that internal $\gamma\gamma$ attenuation
becomes too strong and absorbs the TeV $\gamma$-rays.  In the
particular case of $t_{var} = 1~$day, which is supported by the low
activity and low multifrequency variability observed during the
campaign, $R = 5\times 10^{16}$ cm, that is 2-3 orders of magnitude
larger than the gravitational radius. Under the assumption that the
emission comes from the entire (or large fraction of the)
cross-section of the jet, and assuming a conical jet, the location of
the emitting region would be given by $L \sim R/\theta$, where $\theta
\sim 1/\Gamma \sim 1/\delta$. Therefore, under these assumptions,
which are valid for large distances ($L\gg R_g$) when the outflow is
fully formed, the leptonic model would put the emission region at $L
\sim 10^{3}-10^{4} R_{g}$. We note however that, since the $R$ for the
leptonic model is considered an upper limit on the blob size scale
(see eqn. \ref{Rb_eqn}), this distance should also be considered as an
upper limit as well.

\subsubsection{Particle content and particle acceleration}

The particle content predicted by the hadronic and leptonic scenarios
are different by construction. In the hadronic scenario presented in
\S \ref{hadronicmodelSEDText}, the dominantly radiating particles are
protons, secondary electron/positron pairs, muons, and pions, in
addition to the primary electrons.  In the leptonic scenario, the
dominantly radiating particles are the primary electrons only. In both
cases, the distribution of particles are clearly non-thermal and
acceleration mechanisms are required.

In the leptonic scenario, the PL index $p_1=2.2$, which is the
canonical particle spectral index from efficient 1st-order Fermi
acceleration at the fronts of relativistic shocks, suggests that this
process is at work in Mrk\,421.  For electrons to be picked up by
1st-order Fermi acceleration in perpendicular shocks, their Larmor
radius is required to be significantly larger than the width of the
shock, which for electron-proton plasmas is set by the Larmor radius
of the dynamically dominant particles (electrons or protons). {\bf The
large $\gamma_{min}$ ($= 8 \times 10^{2}$) provided by the model
implies that electrons are efficiently accelerated by the Fermi mechanism  only above this
energy and that below this energy they are accelerated by a different mechanism that produces an extremely hard electron distribution.} Such pre-acceleration mechanisms have been discussed
in the past \citep[e.g.,][]{hos92}. The suggestion  that the Fermi
mechanism picks up only after $\gamma_{min}$ ($= 8 \times 10^{2}$)  suggests a large thickness of the shock, which would
imply that the shock is dominated by (cold) protons. We refer the
reader to the \FermiLAT paper on Mrk\,501 \citep{AbdoMrk501} for more
detailed discussion on this topic.  
%Within this framework, it is required that a pre-acceleration mechanism brings the electrons up to $\gamma \sim 8 \times 10^{2}$.  This pre-acceleration mechanism is required to produce a very hard electron spectrum below $\gamma_{min}$.  Such pre-acceleration mechanisms have been discussed in the past \citep[e.g.,][]{hos92}. 
 In addition, in \S \ref{leptonicmodel} and \S \ref{breaks} we argued that the second
break $\gamma_{brk,2} m_e c^2$ ($\sim 200$ GeV) is probably due to synchrotron
cooling (the electrons radiate most of their energy before existing
the region of size $R$), but the first break $\gamma_{brk,1} m_e c^2$ ($\sim
25$ GeV ) must be related to the acceleration mechanism; and hence the
leptonic model also requires that electrons above the first break are
accelerated less efficiently. At this point it is interesting to note
that the 1-zone SSC model of Mrk\,501 in 2009 (where the source was
also observed mostly in a quiescent state), returned $\gamma_{brk,1} m_e c^2
\sim 20$ GeV with essentially the same spectral change (0.5) in the
electron distribution \citep{AbdoMrk501}.  Therefore, the first break
(presumedly related to the acceleration mechanism) is of the same
magnitude and located approximately at the same energy for both
Mrk\,421 and Mrk\,501, which might suggest a common property in the
quiescent state of HSP BL Lac objects detected at TeV energies.

The presence of intrinsic high energy breaks in the EED electron
energy distribution has been observed in several of the \FermiLAT
blazars \citep[see][]{abd09b, abdoSpectralProperties}. As reported in
\citet{abdoSpectralProperties}, this characteristic was observed on
several FSRQs, and it is present in some LSP-BLLacs, and a small
number of ISP-BLLacs; yet it is absent in all 1LAC HSP-BLLacs.  In
this paper (as well as in \citet{AbdoMrk501}) we claim that such
feature is also present in HSP-BLLacs like Mrk\,421 and Mrk\,501, yet
for those objects, the breaks in the EED can only be accessed through
proper SED modeling because they are smaller in magnitude, and
somewhat smoothed in the high energy component. We note that, for
HSP-BLLacs, the high energy bump is believed to be produced by the EED
upscattering seed-photons from a wide energy range (the synchrotron
photons emitted by the EED itself) and hence all the features from the
EED are smoothed out. On the other hand, in the other blazar objects
like FSRQs, the high energy bump is believed to be produced by the EED
upscattering (external) seed-photons which have a ``relatively
narrow'' energy range. In this later case (external compton), the
features of the EED may be directly seen in the gamma-ray
spectrum. Another interesting observation is that, at least for one of
the FSRQs, 3C\,454.3, the location and the magnitude of the break
seems to be insensitive to flux variations \citep{Ackermann2010}. If
the break observed in Mrk\,421 and Mrk\,501 is of the same nature than
that of 3C\,454.3, we should also expect to see this break at the same
location ($\sim 20$\,GeV) regardless of the activity level of these
sources.

In the hadronic scenario of Figure~\ref{hadronicmodelSED}, the blazar
emission comes from a compact ($R \sim$ a few $R_g$) highly magnetized
emission region, which should be sufficiently far away from the
central engine so that the photon density from the accretion disk is
much smaller than the density of synchrotron photons. The gyroradius
of the highest energy protons ($R_L=\gamma_{p,max} m_p c^2/(e B)$
in Gaussian-cgs units) is $\sim 1.4 \times 10^{14}$ cm, which is a
factor of $\sim$3 times smaller than the radius of the spherical
region responsible for the blazar emission ($R=4 \times 10^{14}$ cm), 
hence (barely) fulfilling the Hillas criterium. The small size of the
emitting region, the ultra-high particle energies and the somewhat 
higher (by factor $\sim$5) particle energy density with respect to 
the magnetic energy density imply that this scenario requires 
extreme acceleration and confinement conditions.

%This scenario might let us speculate that the acceleration of the particles could be
%related to very strong electromagnetic fields generated close to the
%central engine due to differential rotation of the magnetic field
%lines in the inner parts of the accretion disk or due to frame
%dragging in the ergosphere of the rotating super-massive black
%hole \citep[e.g.,][]{acc1,acc2,acc3}.

\subsubsection{Energetics of the jet}

The power of the various components of the flow differ in the two
models. In the SPB model, the particle energy density is about a 
factor of $\sim$5 higher than the magnetic field energy
density and the proton energy density
dominates over that of the electrons by a factor of $\sim 40$. In the
leptonic model the electron energy density dominates over that of the
magnetic field by a factor of 10. By construction, the leptonic model
does not constrain the proton content and hence we need to make
assumptions on the number of protons. It is reasonable to use charge
neutrality to justify a comparable number of electrons and
protons. Under this assumption, the leptonic model predicts that the
energy carried by the electrons (which is dominated by the parameter
$\gamma_{min} \sim 10^{3}$) is comparable to that carried by the
(cold) protons.

The overall jet power determined by the hadronic model is $P_{jet} =
4.4 \times10^{44}$ erg $s^{-1}$.  For the day variability timescale
leptonic model, assuming one cold proton per radiating electron, the
power carried by the protons would be $4.4\times10^{43}$ erg s$^{-1}$,
giving a total jet power of $P_{jet} = 1.9\times 10^{44}$ erg
$s^{-1}$.  In both cases, the computed jet power is a small fraction
($\sim 10^{-2}-10^{-3}$) of the Eddington luminosity for the
supermassive black hole in Mrk\,421 ($2\cdot 10^8 M_\odot$) which is
$L_{Edd}\sim 10^{46}-10^{47}$ erg s$^{-1}$.

\subsection{Interpretation of the Reported Variability}
\label{Mrk421Variability}

In \S\ref{LC} we reported the $\gamma$-ray flux/spectral variations of
Mrk\,421 as measured by the \FermiLAT instrument during the first 1.5
years of operation. The flux and spectral index were determined on
7-day-long time intervals. We showed that, while the $\gamma$-ray flux
above 0.3 GeV flux changed by a factor of $\sim 3$, the PL photon
index variations are consistent with statistical fluctuations (Figure
\ref{fig:Lc7days}) and the spectral variability could only be detected
when comparing the variability in the $\gamma$-ray flux above $2$\,GeV
with that one from the $\gamma$-ray flux below $2$\,GeV. It is worth
pointing out that, in the case of the TeV blazar Mrk\,501, the
$\gamma$-ray flux above $2$\,GeV was also found to vary more than the
$\gamma$-ray flux below $2$\,GeV. Yet unlike Mrk\,421, 
Mrk\,501 was less bright at $\gamma$-rays and the flux
variations above $2$\,GeV seem to be larger, which produced statistically
significant changes in the photon index from the PL fit in the energy
range 0.3-400 GeV \citep[see][]{AbdoMrk501}. In any case, it is
interesting to note that in these two
(classical) TeV objects, the flux variations above few GeV are larger
than the ones below a few GeV, which might suggest that this is a common
property in HSP BL Lac objects detected at TeV energies.

In \S\ref{LC} we also showed (see Figures\ \ref{fig:Lc7daysMW},
\ref{fig:LcFlare_3day}, and \ref{fig:nva}) that the X-ray variability
is significantly higher than that in the $\gamma$-ray band measured by
\FermiLAT. In addition, we also saw that the $15-50$\,keV (BAT) and
the $2-10$\,keV (ASM) fluxes are positively correlated, and that the
BAT flux is more variable than the ASM flux. In other words, when the
source flares in X-rays, the X-ray spectrum becomes harder.

\begin{figure}[t]
  \centering
  \includegraphics[scale=0.45]{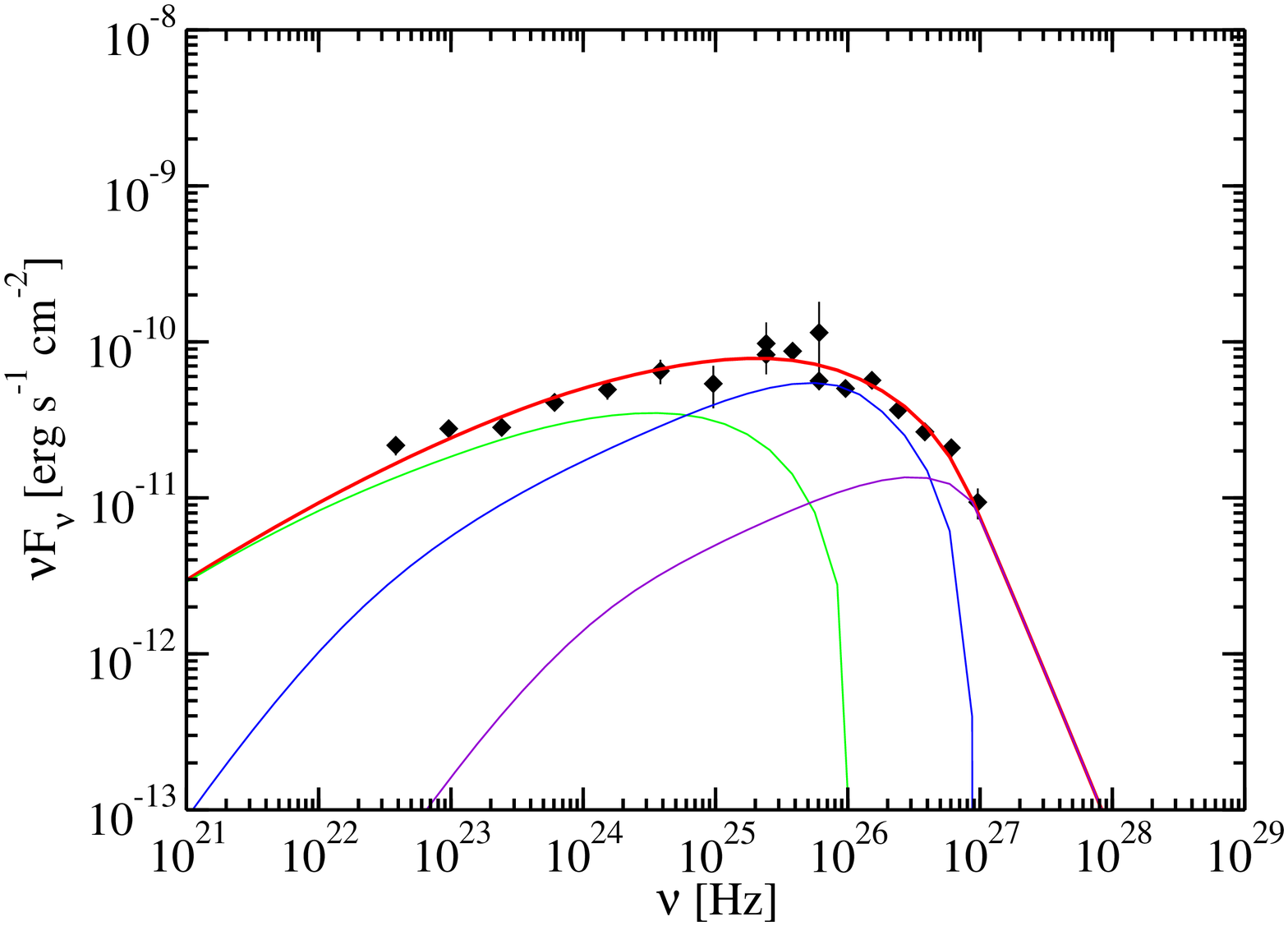}
  \includegraphics[scale=0.45]{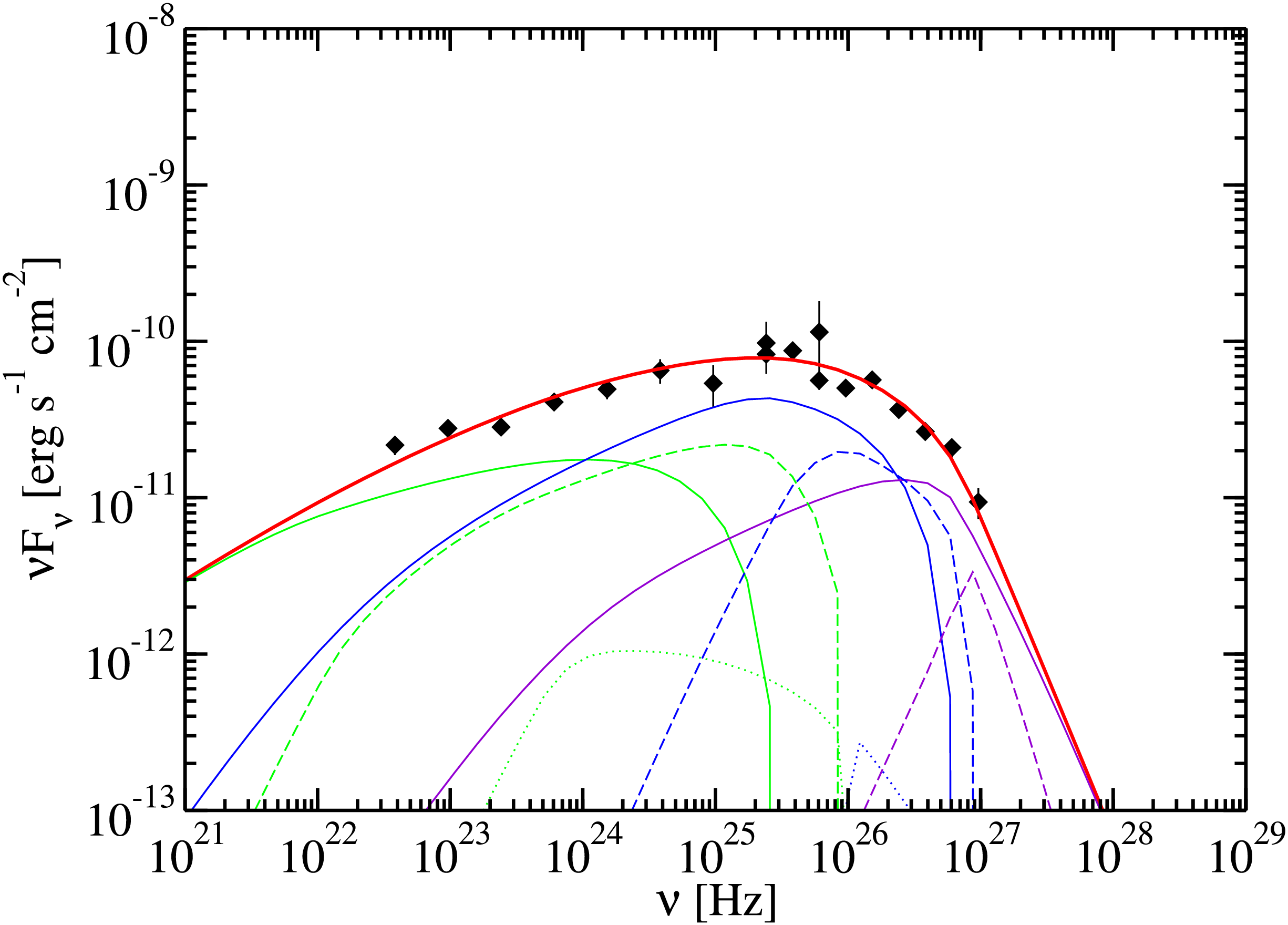}
  \caption{Decomposition of the high energy bump of the SSC continuum for Mrk\,421. The data points are the same as in the high energy bump from  Figure~\ref{modelSED}. The SSC fit to the average spectrum is denoted by the red solid curve. {\bf Top:} Contributions of the different segments of electrons comptonizing the whole synchrotron continuum (green curve: $\gamma_{min} < \gamma < \gamma_{br,\,1}$; blue curve: $\gamma_{br,\,1} < \gamma < \gamma_{br,\,2}$; purple curve: $\gamma_{br,\,2} < \gamma$). {\bf Bottom:} Contributions of the different segments of electrons (as in the top panel) Comptonizing different segments of the synchrotron continuum (solid curves: $\nu < \nu_{br,\,1} \simeq 5.3 \times 10^{15}$\,Hz; dashed curves: $\nu_{br,\,1} < \nu < \nu_{br,\,2} \simeq 1.3 \times 10^{17}$\,Hz; dotted curves, corresponding to $\nu > \nu_{br,\,2}$)}
  \label{fig:SED_GammaZoom}
 \end{figure}

In order to understand this long baseline X-ray/$\gamma$-ray
variability within our leptonic scenario, we decomposed the
$\gamma$-ray bump of the SED into the various contributions from the
various segments of the EED, according to our 1-zone SSC model, in a
similar way as it was done in \cite{Tavecchio1998}. This
is depicted in Figure~\ref{fig:SED_GammaZoom}. The contributions of
different segments of the EED are indicated by different colors. As
shown, the low-energy electrons, $\gamma_{min} \leq \gamma <
\gamma_{br,\,1}$, which are emitting synchrotron photons up to the
observed frequencies of $\simeq 5.2 \times 10^{15}$\,Hz, dominate the
production of $\gamma$-rays up to the observed photon energies of
$\sim$20\,GeV (green line). The contribution of higher energy
electrons with Lorentz factors $\gamma_{br,\,1} \leq \gamma <
\gamma_{br,\,2}$ is pronounced within the observed synchrotron range
$5 \times 10^{15}- 10^{17}$\,Hz, and at $\gamma$-ray energies from
$\sim$20\,GeV up to $\sim$\,TeV (blue line). Finally, the highest
energy tail of the electron energy distribution, $\gamma \geq
\gamma_{br,\,2}$, responsible for the production of the observed X-ray
synchrotron continuum ($>0.5$\,keV) 
generates the bulk of $\gamma$-rays with the observed energies
$>$\,TeV (purple line). Because of the electrons upscattering the
broad energy range of synchrotron photons, the emission of the
different electron segments are somewhat connected, as shown in the
bottom plot of Figure~\ref{fig:SED_GammaZoom}. Specifically, the low
energy electrons have also contributed to the TeV photon flux through
the emitted synchrotron photons which are being upscattered by the
high energy electrons. Hence, changes in the number of low energy
electrons should also have an impact on the TeV photon flux. However,
note that the synchrotron photons emitted by the high energy
electrons, which are up-scattered in the Klein-Nishina regime, do not
have any significant contribution to the gamma-ray flux, thus
changes in the number of high energy electrons (say $\gamma >
\gamma_{br,\,2}$) will not significantly change the MeV/GeV photon flux.

Within our 1-zone SSC scenario, the $\gamma$-rays measured by
\FermiLAT are mostly produced by the low energy electrons ($\gamma
\leq \gamma_{br,\,1}$) while the X-rays seen by ASM and BAT are mostly
produced by the highest energy electrons ($\gamma \geq
\gamma_{br,\,2}$). In this scenario, the significantly higher
variability in the X-rays with respect to that of
$\gamma$-rays suggests that the flux variations in Mrk\,421 are
dominated by changes in the number of the highest energy
electrons. Note that the same trend is observed in the X-rays (ASM
versus BAT) and $\gamma$-rays (below versus above $2$\,GeV); the
variability in the emission increases with the energy of the radiating
electrons.

The greater variability in the radiation produced by the highest
energy electrons is not surprising.  The cooling timescales of the
electrons from synchrotron and inverse Compton (in the Thomson regime) losses
scale as $t\propto\gamma^{-1}$, and hence it is expected that the
emission from higher energy electrons will be the most variable.
However, since the high energy electrons are the ones losing their
energy fastest, in order to keep the source emitting in X-rays,
injection (acceleration) of electrons up to the highest energies is
needed. This injection (acceleration) of high energy electrons could
well be the origin of the flux variations in Mrk\,421. The details of
this high energy electron injection could be parameterized by changes
in the parameters $\gamma_{br,\,2}$, $p_3$ and $\gamma_{max}$ within
the framework of the 1-zone SSC model that could result from episodic
acceleration events \citep{Perlman05}. The characterization of the SED
evolution (and hence SSC parameter variations) will be one of the
prime subjects of the forthcoming publications with the
multi-instrument variability and correlation during the campaigns in
2009\footnote{ For details of the 2009 campaign, see the URL \\
{\scriptsize
\url{https://confluence.slac.stanford.edu/display/GLAMCOG/Campaign+on+Mrk421+(Jan+2009+to+May+2009)}
}} and 2010\footnote{ For details of the 2010 campaign, see the URL \\
{\scriptsize
\mbox{\texttt{https://confluence.slac.stanford.edu/display/GLAMCOG/Campaign+on+Mrk421+\%28December+2009+to+December+2010\%29}}}}. We
note here that SSC models, both one-zone and multi-zone
\citep[e.g.][]{graff08}, predict a positive correlation between the X-rays and
the TeV $\gamma$-rays measured by IACTs. Indeed, during the 2010 campaign
the source was detected in a flaring state with the TeV instruments
(see ATel \#2443). Such an X-ray/TeV correlation has been established
in the past for this object \citep[see][]{Maraschi1999}, although the relation is not simple.
Sometimes it is linear and other times quadratic
\citep[e.g.][]{fossati08}. The complexity of this correlation is also
consistent with our 1-zone SSC model; the X-rays are produced by
electrons with $\gamma > \gamma_{br,\,2}$, while the TeV photons are
produced by electrons with $\gamma > \gamma_{br,\,1}$, and is 
indirectly affected by the electrons with $\gamma < \gamma_{br,\,1}$
through the emitted synchrotron photons that are used as seed photons
for the inverse Compton scattering (see the bottom plot of
Figure~\ref{fig:SED_GammaZoom}).

We also note that the 1-zone SSC scenario presented here predicts a
direct correlation on the basis of simultaneous data sets between the
low energy gamma-rays (from {\em Fermi}) and the sub-millimeter (SMA)
and optical frequencies, since both energy bands are produced by the
lowest energy electrons in the source. On the other hand, our SPB
model fit does not require such a strict correlation, but there could
be a loose correlation if electrons and protons are accelerated
together. In particular, a direct correlation with zero time lag
between the mm radio frequencies and the $\gamma$-rays is not expected
in our SPB model because the radiation at these two energy bands are
produced at different sites.  The radiation in the X- and $\gamma$-ray
band originates from the primary electrons, and from {\bf the protons and} secondary
particles created by proton-initiated processes,
respectively. Consequently, although a loose correlation between the
X-ray and $\gamma$-ray band can be expected if protons and electrons
are accelerated together, a strict correlation with zero time lag is
rather unlikely in our model fit.

During the 2009 and 2010 campaigns, Mrk\,421 was very densely sampled
during a very long baseline (4.5 and 6 months for the 2009 and 2010
campaign, respectively) and hence these data sets will provide
excellent information for performing a very detailed study of these
multi-band relations. In particular, during the campaign in 2010,
there were regular observations with VLBA and SMA, which will allow us
to study with a greater level of detail the relationship between the
rising parts of the low energy and high energy bumps, where the
predictions from the leptonic and hadronic models differ.

%% ============================================================================
%%
%% SECTION 8 -- Conclusion 
%%
%% ============================================================================

\section{Conclusions} 
\label{Conclusions}

In this work, we reported on the $\gamma$-ray activity of Mrk\,421 as
measured by the LAT instrument on board the \Fermi satellite during
its first 1.5 years of operation, from 2008 August 5 (MJD 54683) to
2009 March 12 (MJD 55248).  Because of the large leap in capabilities
of LAT with respect to its predecessor, EGRET, this is the most
extensive study of the $\gamma$-ray activity of this object at GeV
photon energies to date. The \FermiLAT spectrum (quantified with a
single power-law function) was evaluated for 7-day-long time
intervals. The average photon flux above $0.3$\,GeV was found to be
$(7.23 \pm 0.16) \times 10^{-8}$\,ph\,cm$^{-2}$\,s$^{-1}$, and the
average photon index $1.78 \pm 0.02$. The photon flux changed
significantly (up to a factor $\sim 3$) while the
spectral variations were mild. The variations in the PL photon index
were not statistically significant, yet the light curves and
variability quantification below and above $2$\,GeV showed that the
high $\gamma$-ray energies vary more than the low energy
$\gamma$-rays.  We found $F_{var} (E<2$\,GeV$) = 0.16 \pm 0.04$ while $F_{var}
(E>2$\,GeV$) = 0.33 \pm 0.04$. We compared the LAT $\gamma$-ray activity
in these two energy ranges ($0.2-2$\,GeV and $> 2$\,GeV) with the X-ray
activity recorded by the all-sky instruments \RXTEc/ASM ($2-10$\,keV)
and \Swiftc/BAT ($15-50$\,keV). We found that X-rays are significantly
more variable than $\gamma$-rays, with no significant (\lapp $2
\sigma$) correlation between them. We also found that, within the
X-ray and $\gamma$-ray energy bands, the variability increased with
photon energy.  The physical interpretation of this result within the
context of the 1-zone SSC model is that the variability in the
radiation increases with the energy of the electrons that produce
them, which is expected given the radiating time scales
for synchrotron and inverse Compton emission.

We also presented the first results from the
4.5-month-long multifrequency campaign on Mrk\,421, which lasted from
2009 January 19 (MJD 54850) to 2009 June 1 (MJD 54983). During this 
time period, the source was systematically observed from radio
to TeV energies. 
%Here, we only focused on the
%average SED emerging from the campaign, leaving any further studies on
%the multifrequency  variability and correlations for a forthcoming
%publication. 
Because of the low activity and low variability shown
during this campaign, the compiled data provided us with the best
SED yet of Mrk\,421 in the low/quiescent state.

The broadband SED was modeled with two different scenarios: a leptonic
(1 zone SSC) and a hadronic model (SPB). Both frameworks are able to
describe reasonably well the average SED, implying comparable powers
for the jet emission, which constitute only a small fraction
($\sim10^{-2}-10^{-3}$) of the Eddington luminosity.  However, those
models differ on the predicted environment for the blazar emission: 
the leptonic scenario constrains the size to be $R\la 10^4~R_g$,
the magnetic field to $B\sim 0.05$ G and particles (electrons) with
energies up to $\sim 5 \cdot10^{13}$ eV while, if $\alpha_e=\alpha_p$, 
our hadronic scenario implies a size of the
emitting region of a few $R_g$, a magnetic field $B\sim 50$ G and
particles (protons) with energies up to $\sim 2 \cdot10^{18}$ eV,
which requires extreme conditions for particle acceleration and 
confinement.

The leptonic scenario suggests that the acceleration of the radiating
particles (electrons) is through diffusive shock acceleration in
relativistic shocks mediated by cold protons, and that this mechanism
accelerates particles (electrons) less efficiently above an energy of
$\sim 25$ GeV, which is comparable to what was reported in
\cite{AbdoMrk501} for another classical TeV blazar, Mrk\,501.  In
addition, unlike what was observed for Mrk\,501, in the case of
Mrk\,421 a stronger-than-canonical electron cooling break was required
to reproduce the observed SED, which might suggest that the 
blazar emitting region is inhomogeneous.

Within the SSC model (Figure~\ref{modelSED}), the observed
X-ray/$\gamma$-ray variability during the first 1.5 years of \Fermi
operation indicates that the flux variations in Mrk\,421 are produced
by acceleration of the highest energy electrons, which radiate in the
X-ray and TeV bands, and lose energy, radiating as they do so in the
optical and GeV range. In our hadronic model
(Figure~\ref{hadronicmodelSED}), a rather loose correlation between
the X- and $\gamma$-ray band is expected if electrons and protons are
accelerated together. A forthcoming publication will report on whether
these emission models can reproduce the multi-band flux variations
observed during the intensive campaigns on Mrk\,421 performed in 2009
and 2010. Those studies should help us distinguish between the
hadronic and the leptonic scenarios and eventually lead to a better
understanding of one of the fundamental mysteries of blazars: how flux
variations are produced.

%% ============================================================================
%%
%% ACKNOWLEDGEMENTS
%%
%% ============================================================================

%{\large Acknowledgments}

\acknowledgments

The authors of the paper thank the anonymous referee for very well
organized and constructive comments that helped improving the quality
and clarity of this publication.

The \FermiLAT Collaboration acknowledges the generous support of a number of agencies and institutes that have supported the \FermiLAT Collaboration. These include the National Aeronautics and Space Administration and the Department of Energy in the United States, the Commissariat \`a l'Energie Atomique and the Centre National de la Recherche Scientifique / Institut National de Physique Nucl\'eaire et de Physique des Particules in France, the Agenzia Spaziale Italiana and the Istituto Nazionale di Fisica Nucleare in Italy, the Ministry of Education, Culture, Sports, Science and Technology (MEXT), High Energy Accelerator Research Organization (KEK) and Japan Aerospace Exploration Agency (JAXA) in Japan, and the K.\ A.\ Wallenberg Foundation, the Swedish Research Council and the Swedish National Space Board in Sweden. Additional support for science analysis during the operations phase is gratefully
acknowledged from the Istituto Nazionale di Astrofisica in Italy and the Centre National d'\'Etudes Spatiales in France.

The MAGIC collaboration would like to thank the Instituto de Astrof\'{\i}sica de Canarias for the excellent working conditions at the Observatorio del Roque de los Muchachos in La Palma.
The support of the German BMBF and MPG, the Italian INFN,  the Swiss National Fund SNF, and the Spanish MICINN is gratefully acknowledged. This work was also supported by the Marie Curie program, by the CPAN CSD2007-00042 and MultiDark CSD2009-00064 projects of the Spanish Consolider-Ingenio 2010 programme, by grant DO02-353 of the Bulgarian NSF, by grant 127740 of 
the Academy of Finland, by the YIP of the Helmholtz Gemeinschaft, by the DFG Cluster of Excellence ``Origin and Structure of the Universe'', and by the Polish MNiSzW Grant N N203 390834.

We acknowledge the use of public data from the \Swift and \RXTE data
archive. The Mets\"ahovi team acknowledges the support from the
Academy of Finland to the observing projects (numbers 212656, 210338,
among others).  This research has made use of data obtained from the
National Radio Astronomy Observatory's Very Long Baseline Array
(VLBA), projects BK150, BP143 and BL149 (MOJAVE). The National Radio Astronomy
Observatory is a facility of the National Science Foundation operated
under cooperative agreement by Associated Universities, Inc.
St.Petersburg University team acknowledges support from Russian RFBR
foundation via grant 09-02-00092. AZT-24 observations are made within
an agreement between  Pulkovo, Rome and Teramo observatories. This
research is partly based on observations with the 100-m telescope of
the MPIfR (Max-Planck-Institut f\"ur Radioastronomie) at Effelsberg,
as well as with the Medicina and Noto telescopes operated by INAF -
Istituto di Radioastronomia. RATAN-600 observations were supported in part by the RFBR grant
08-02-00545 and the OVRO 40 m program was funded in part by NASA (NNX08AW31G) and the NSF (AST-0808050).
The Submillimeter Array is a joint
project between the Smithsonian  Astrophysical Observatory and the
Academia Sinica Institute of Astronomy and  Astrophysics and is funded
by the Smithsonian Institution and the Academia Sinica. M. Villata
organized the optical-to-radio observations by GASP-WEBT as the
president of the collaboration. Abastumani Observatory team
acknowledges financial support by the Georgian National Science
Foundation through grant GNSF/ST07/4-180.

\end{document}